\documentclass[journal,10pt]{IEEEtran}

\usepackage{amssymb,amsmath,amsthm, mathtools}
\usepackage{graphicx}
\usepackage{cite}
\usepackage{setspace}
\usepackage{latexsym}
\usepackage{verbatim}
\usepackage{amsmath}
\usepackage{amssymb}
\usepackage{subfigure}
\usepackage{url}
\usepackage{epstopdf}
\usepackage{algorithm}
\usepackage{times}
\usepackage{stfloats}
\usepackage{multirow}
\usepackage{enumerate}
\usepackage{xspace}
\usepackage{longtable}
\usepackage{mathrsfs}
\usepackage{mathtools}
\usepackage{algcompatible}
\usepackage[noend]{algpseudocode}
\usepackage{amssymb}
\usepackage{pifont}
\usepackage{caption}
\usepackage{multicol}
\usepackage{color}
\usepackage[table]{xcolor}
\usepackage{longtable}

\newcommand{\PPSS}{\ensuremath {\mathit{PPSS}}{\xspace}}
\newcommand{\fc}{\ensuremath {\mathit{FC}}{\xspace}}

\newcommand{\crn}{\ensuremath {\mathit{CRN}}{\xspace}}

\newcommand{\sk}{\ensuremath {\mathit{sk}}{\xspace}}
\newcommand{\pk}{\ensuremath {\mathit{pk}}{\xspace}}
\newcommand{\skhors}{\ensuremath {\mathit{SK_{DS}}}{\xspace}}
\newcommand{\pkhors}{\ensuremath {\mathit{PK_{DS}}}{\xspace}}
\newcommand{\su}{\ensuremath {\mathit{SU}}{\xspace}}

\newcommand{\sign}{\ensuremath {\mathit{\sigma}}{\xspace}}
\newcommand{\decision}{\ensuremath {\mathit{dec}}{\xspace}}
\newcommand{\loDec}{\ensuremath {\mathit{D}}{\xspace}}
\newcommand{\loHyp}{\ensuremath {\mathit{h}}{\xspace}}
\newcommand{\Hyp}{\ensuremath \mathcal{H}{\xspace}}
\newcommand{\pu}{\ensuremath {\mathit{PU}}{\xspace}}
\newcommand{\gw}{\ensuremath {\mathit{GW}}{\xspace}}
\newcommand{\rss}{\ensuremath {\mathit{RSS}}{\xspace}}
\newcommand{\ta}{\ensuremath {\mathit{\tau}}{\xspace}}
\newcommand{\lam}{\ensuremath {\mathit{\lambda}}{\xspace}}
\newcommand{\ope}{\ensuremath {\mathit{OPE}}{\xspace}}
\newcommand{\key}{\ensuremath {\mathit{k}}{\xspace}}
\newcommand{\ciphFC}{\ensuremath {\mathit{\theta}}{\xspace}}
\newcommand{\ciphU}{\ensuremath {\mathit{\varsigma}}{\xspace}}
\newcommand{\rsa}{\ensuremath {\mathit{N}}{\xspace}}
\newcommand{\ciphGW}{\ensuremath {\mathit{\zeta}}{\xspace}}
\newcommand{\nbr}{\ensuremath {\mathit{n}}{\xspace}} 
\newcommand{\gam}{\ensuremath {\mathit{\gamma}}{\xspace}} 
\newcommand{\bin}{\ensuremath {\mathit{b}}{\xspace}} 
\newcommand{\votes}{\ensuremath {\mathit{v}}{\xspace}} 
\newcommand{\chn}{\ensuremath {\mathit{chn}}{\xspace}}
\newcommand{\pol}{\ensuremath {\mathit{Pol}}{\xspace}} 
\newcommand{\sr}{\ensuremath {\mathit{\mu}}{\xspace}}
\newcommand{\chr}{\ensuremath {\mathit{\beta}}{\xspace}}

\newcommand{\kap}{\ensuremath {\mathit{\kappa}}{\xspace}}\newcommand{\posr}{\ensuremath {\mathit{\varrho}}{\xspace}}
\newcommand{\ran}{\ensuremath {\mathit{R}}{\xspace}}
\newcommand{\alhv}{\ensuremath {\mathit{\alpha}}{\xspace}} 
\newcommand{\ECDLP}{\ensuremath {\mathit{ECDLP}}{\xspace}}
\newcommand{\SCPU}{\ensuremath {\mathit{SCPU}}{\xspace}}
\newcommand{\TGECDH}{\ensuremath {\mathit{TGECDH}}{\xspace}}
\newcommand{\POLLAM}{{\em Pollard-Lambda}{\xspace}}
\newcommand{\ECELG}{\ensuremath {\mathit{ECElG}}{\xspace}}

\newcommand{\ym}{\ensuremath {\mathit{YM}}{\xspace}}
\newcommand{\yme}{\ensuremath {\mathit{YM.ECElG}}{\xspace}}
\newcommand{\ymg}{\ensuremath {\mathit{YM.ElGamal}}{\xspace}}
\newcommand{\aes}{\ensuremath {\mathit{AES}}{\xspace}} 
\newcommand{\blck}{\ensuremath {\mathit{\epsilon_{\En}}}{\xspace}} 
\newcommand{\pr}{\ensuremath {\mathit{\pi}}{\xspace}}
\newcommand{\OEnc}[2]{\ensuremath{\mathit{OPE}.\mathcal{E}_{#1}\mskip-1mu(#2)}}
\newcommand{\OEncc}{\ensuremath{\mathit{OPE}.\mathcal{E}}}
\newcommand{\LU}{\ensuremath{\mathcal{L}_1}}
\newcommand{\LF}{\ensuremath{\mathcal{L}_2}}
\newcommand{\LG}{\ensuremath{\mathcal{L}_3}}
\newcommand{\V}{\ensuremath{\vec{V}}}
\newcommand{\Z}{\ensuremath{\vec{Z}}}
\newcommand{\A}{\ensuremath{\mathcal{A}}}

\newcommand{\Kb}{\ensuremath{\bar{K}}}
\newcommand{\as}{\ensuremath {\leftarrow}{\xspace}}

\newcommand{\En}{\ensuremath{\mathcal{E}}{\xspace}}
\newcommand{\De}{\ensuremath{\mathcal{D}}{\xspace}}
\newcommand{\Enc}[2]{\ensuremath{\mathcal{E}_{#1}\mskip-1mu(#2)}}
\newcommand{\Dec}[2]{\ensuremath{\mathcal{D}_{#1}\mskip-1mu(#2)}}
\newcommand{\ECEG}{\ensuremath {\mathit{ECEG}}{\xspace}}
\newcommand{\HORS}{\ensuremath {\mathit{HORS}}{\xspace}}
\newcommand{\LPOS}{\ensuremath {\mathit{LPOS}}{\xspace}}

\newcommand{\SGN}{\ensuremath {\mathit{SGN}}{\xspace}}
\newcommand{\ROLPOS}{{\em LP-2PSS}{\xspace}}   
\newcommand{\PDAFT}{\ensuremath {\mathit{PDAFT}}{\xspace}}
\newcommand{\LPGW}{{\em LP-3PSS}{\xspace}}   

\newcommand{\srv}{\ensuremath {\mathit{y}}{\xspace}}
\newcommand{\algrule}[1][.2pt]{\par\vskip.5\baselineskip\hrule height #1\par\vskip.5\baselineskip}

\newtheorem{myremark}{Remark}{\bfseries}{\rmfamily}
\newtheorem{definition}{Definition}{\bfseries}{\rmfamily}
\newtheorem{mytheorem}{Theorem}{\bfseries}{\rmfamily}
{\bfseries}{\rmfamily}
\newtheorem{mycorollary}{Corollary}{\bfseries}{\rmfamily}
\newtheorem{fact}{Fact}{\bfseries}{\rmfamily}
\newtheorem{assumption}{Security Assumptions}{\bfseries}{\rmfamily}
\newtheorem{objective}{Security Objectives}{\bfseries}{\rmfamily}
{\bfseries}{\rmfamily}
\makeatletter
\newcommand*{\da@rightarrow}{\mathchar"0\hexnumber@\symAMSa 4B }
\newcommand*{\da@leftarrow}{\mathchar"0\hexnumber@\symAMSa 4C }
\newcommand*{\xdashrightarrow}[2][]{%
  \mathrel{%
    \mathpalette{\da@xarrow{#1}{#2}{}\da@rightarrow{\,}{}}{}%
  }%
}
\newcommand{\xdashleftarrow}[2][]{%
  \mathrel{%
    \mathpalette{\da@xarrow{#1}{#2}\da@leftarrow{}{}{\,}}{}%
  }%
}
\newcommand*{\da@xarrow}[7]{%
  \sbox0{$\ifx#7\scriptstyle\scriptscriptstyle\else\scriptstyle\fi#5#1#6\m@th$}%
  \sbox2{$\ifx#7\scriptstyle\scriptscriptstyle\else\scriptstyle\fi#5#2#6\m@th$}%
  \sbox4{$#7\dabar@\m@th$}%
  \dimen@=\wd0 %
  \ifdim\wd2 >\dimen@
    \dimen@=\wd2 %
  \fi
  \count@=2 %
  \def\da@bars{\dabar@\dabar@}%
  \@whiledim\count@\wd4<\dimen@\do{%
    \advance\count@\@ne
    \expandafter\def\expandafter\da@bars\expandafter{%
      \da@bars
      \dabar@ 
    }%
  }%
  \mathrel{#3}%
  \mathrel{%
    \mathop{\da@bars}\limits
    \ifx\\#1\\%
    \else
      _{\copy0}%
    \fi
    \ifx\\#2\\%
    \else
      ^{\copy2}%
    \fi
  }%
  \mathrel{#4}%
}
\makeatother
\begin{document}
%
\title{Preserving the Location Privacy of Secondary Users in Cooperative Spectrum Sensing}
%
%
%

\author{Mohamed~Grissa,~\IEEEmembership{Student~Member,~IEEE,}
        Attila~A.~Yavuz,~\IEEEmembership{Member,~IEEE,}
        and~Bechir~Hamdaoui,~\IEEEmembership{Senior~Member,~IEEE}
    
\thanks{This work was supported in part by the US National Science Foundation under NSF award CNS-1162296. Mohamed Grissa, Attila~A.~Yavuz and~Bechir~Hamdaoui are with the Electrical Engineering and Computer Science (EECS) Department, Oregon State University, Corvallis, OR 97331-5501, USA (e-mail: grissam,attila.yavuz,hamdaoui@oregonstate.edu).}
\thanks{\copyright~2016 IEEE. Personal use of this material is permitted. Permission from IEEE must be obtained for all other uses, in any current or future media, including reprinting/republishing this material for advertising or promotional purposes, creating new collective works, for resale or redistribution to servers or lists, or reuse of any copyrighted component of this work in other works.}


}

\maketitle

\begin{abstract}
Cooperative spectrum sensing, despite its effectiveness in enabling dynamic spectrum access, suffers from location privacy threats, merely because Secondary Users (\su s)' sensing reports that need to be shared with a fusion center to make spectrum availability decisions are highly correlated to the users' locations. It is therefore important that cooperative spectrum sensing schemes be empowered with privacy preserving capabilities so as to provide \su s with incentives for participating in the sensing task. In this paper, we propose privacy preserving protocols that make use of various cryptographic mechanisms to preserve the location privacy of \su s while performing reliable and efficient spectrum sensing.
We also present cost-performance tradeoffs. The first consists on using an additional architectural entity at the benefit of incurring lower computation overhead by relying only on symmetric cryptography. The second consists on using an additional secure comparison protocol at the benefit of incurring lesser architectural cost by not requiring extra entities. Our schemes can also adapt to the case of a malicious Fusion Center (\fc) as we discuss in this paper. We also show that not only are our proposed schemes secure and more efficient than existing alternatives, but also achieve fault tolerance and are robust against sporadic network topological changes.
\end{abstract}

\begin{IEEEkeywords}
Location privacy, secure cooperative spectrum sensing, order preserving encryption, cognitive radio networks.
\end{IEEEkeywords}

\IEEEpeerreviewmaketitle

\section{Introduction}
\label{sec:Introduction}

\IEEEPARstart{C}{}ooperative spectrum sensing is a key component of cognitive radio networks ($\crn$s) essential for enabling dynamic and opportunistic spectrum access~\cite{akyildiz2011,guizani2015large,khalfi2015distributed}. It consists of having secondary users (\su s) sense the licensed channels on a regular basis and collaboratively decide whether a channel is available prior to using it so as to avoid harming primary users (\pu s).  %
One of the most popular spectrum sensing techniques is energy detection, thanks to its simplicity and ease of implementation, which essentially detects the presence of \pu's signal by measuring and relying on the energy strength of the sensed signal, commonly known as the received signal strength (\rss)~\cite{fatemieh2011using}.
{\let\thefootnote\relax\footnote{{\\Digital Object Identifier 10.1109/TIFS.2016.2622000}}}

Broadly speaking, cooperative spectrum sensing techniques can be classified into two categories: centralized and distributed~\cite{akyildiz2011}.
In centralized techniques, a central entity called fusion center (\fc) orchestrates the sensing operations as follows. It selects one channel for sensing and, through a control channel, requests that each \su~perform local sensing on that channel and send its sensing report (e.g., the observed \rss~value) back to it. It then combines the received sensing reports, makes a decision about the channel availability, and diffuses the decision back to the \su s. In distributed sensing techniques, \su s do not rely on a \fc~for making channel availability decisions. They instead exchange sensing information among one another to come to a unified decision~\cite{akyildiz2011}.

Despite its usefulness and effectiveness in promoting dynamic spectrum access, cooperative spectrum sensing suffers from serious security and privacy threats. One big threat to \su s, which we tackle in this work, is location privacy, which can easily be compromised due to the wireless nature of the signals communicated by \su s during the cooperative sensing process. In fact, it has been shown that \rss~values of $\su$s are highly correlated to their physical locations~\cite{li2012location}, thus making it easy to compromise the location privacy of \su s when sending out their sensing reports.
The fine-grained location, when combined with other publicly available information, could easily be exploited to infer private information about users~\cite{wicker2012loss}. Examples of such private information are shopping patterns, user preferences, and user beliefs, just to name a few~\cite{wicker2012loss}. With such privacy threats and concerns, $\su$s may refuse to participate in the cooperative sensing tasks.
It is therefore imperative that cooperative sensing schemes be enabled with privacy preserving capabilities that protect the location privacy of \su s, thereby encouraging them to participate in such a key $\crn$ function, the spectrum sensing.

In this paper, we propose two efficient privacy-preserving schemes with several variants for cooperative spectrum sensing. These schemes exploit various cryptographic mechanisms to preserve the location privacy of \su s while performing the cooperative sensing task reliably and efficiently.

In addition, we study the cost-performance tradeoffs of the proposed schemes, and show that higher privacy and better performance can be achieved, but at the cost of deploying an additional architectural entity in the system.
We show that our proposed schemes are secure and more efficient than their existing counterparts, and are robust against sporadic topological changes and network dynamism.

\subsection{Related Work}
\label{sec:related}
Security and privacy in $\crn$s have gained some attention recently. Adem et al.~\cite{adem2016mitigating} addressed jamming attacks in \crn s. Yan et al.~\cite{6195839} discussed security issues in fully distributed cooperative sensing. Qin et al.\cite{qin2014preserving} proposed a privacy-preserving protocol for \crn~transactions using a commitment scheme and zero-knowledge proof. Wang et al.~\cite{wang2015privacy} proposed a privacy preserving framework for collaborative spectrum sensing in the context of multiple service providers.

Location privacy, though well studied in the context of location-based services (LBS)~\cite{6567111,6567112}, has received little attention in the context of \crn s~\cite{gao2013location,liu2013location,li2012location}. Some works focused on location privacy but not in the context of cooperative spectrum sensing (e.g., database-driven spectrum sensing~\cite{grissa2015cuckoo,gao2013location}  and dynamic spectrum auction~\cite{liu2013location}) and are skipped since they are not within this paper's scope.

In the context of cooperative spectrum sensing, Shuai et al.~\cite{li2012location} showed that \su s' locations can easily be inferred from their \rss~reports, and called this the SRLP (single report location privacy) attack. They also identified the DLP (differential location privacy) attack, where a malicious entity can estimate the \rss~(and hence the location) of a leaving/joining user from the variations in the final aggregated \rss~measurements before and after user's joining/leaving of the network. They finally proposed \PPSS~to address these two attacks. Despite its merits, \PPSS~has several limitations: (i) It needs to collect all the sensing reports in order to decode the aggregated result. This is not fault tolerant, since some reports may be missing due, for example, to the unreliable nature of wireless channels. (ii) It cannot handle dynamism if multiple users join or leave the network simultaneously. (iii) The pairwise secret sharing requirement incurs extra communication overhead and delay. (iv) The underlying encryption scheme requires solving the {\em Discrete Logarithm Problem}, which is possible only for very small plaintext space and can be extremely costly.
Chen et al.~\cite{chen2014pdaft} proposed \PDAFT, a fault-tolerant and privacy-preserving data aggregation scheme for smart grid communications. \PDAFT, though proposed in the context of smart grids, is suitable for cooperative sensing schemes. But unlike \PPSS, \PDAFT~relies on an additional semi-trusted entity, called gateway, and like other aggregation based methods, is prone to the DLP attack. In our previous work \cite{grissa2015location} we proposed an efficient scheme called \LPOS~to overcome the limitations that existent approaches suffer from. \LPOS~combines order preserving encryption and yao's millionaire protocol to provide a high location privacy to the users while enabling an efficient sensing performance.

\subsection{Our Contribution} \label{subsec:OurContribution}
In this paper, we propose two location privacy-preserving schemes for cooperative spectrum sensing that achieve:

\begin{itemize}
\item Location privacy of secondary users while performing the cooperative spectrum sensing effectively and reliably.

\item Fault tolerance and robustness against network dynamism (e.g., multiple \su s join/leave the network) and failures (e.g., missed sensing reports).

\item Reliability and resiliency against malicious users via an efficient reputation mechanism.

\item Accurate spectrum availability decisions via half-voting rules while incurring minimum communication and computation overhead.
\end{itemize}

Compared to our preliminary works~\cite{grissa2016an} and~\cite{grissa2015location}, this paper provides a more efficient version of \LPOS~\cite{grissa2015location}, referred to as \ROLPOS~in this paper, that is also robust against malicious users and adapted to a stronger threat model for \fc. Besides, this paper provides another variant of \LPGW~\cite{grissa2016an} that improves the crytpographic end-to-end delay. Finally, this paper provides an improved security analysis and more comprehensive performance analysis.

The reason why we present two variants is to give more options to system designers to decide which topology and which approach is more suitable to their specific requirements. There are tradeoffs between the two options. While \ROLPOS~provides location privacy guarantees without needing to introduce an extra architectural entity, it requires relatively high computational overhead due to the use of the Yao's millionaires' protocol. On the other hand, \LPGW~provides stronger privacy guarantees (as the private inputs are shared among 3 non-colluding entities) and reduces the computational overhead substantially when compared to \ROLPOS, but at the cost of introducing an extra architectural entity.

The remainder of this paper is organized as follows. Section~\ref{sec:system} provides our system and security threat models. Section~\ref{sec:Preliminaries} presents our preliminary concepts and definitions. Section~\ref{sec:replpos} and~\ref{sec:lpgw} provide an extensive explanation of the proposed schemes. Section~\ref{sec:SecAnalysis} gives the security analysis of these schemes. Section~\ref{sec:PerformanceAnalysis} presents their performance analysis and a comparison with existent approaches. Finally, Section~\ref{sec:Conclusion} concludes this work.

\section{System and Security Threat Models}\renewcommand{\figurename}{Fig.}
\label{sec:system}

\subsection{System Model}
We consider a cooperative spectrum sensing architecture that consists of a \fc~and a set of \su s.

Each \su~is assumed to be capable of measuring \rss~on any channel by means of an energy detection method~\cite{fatemieh2011using}.
In this cooperative sensing architecture, the \fc~combines the sensing observations collected from the \su s, decides about the spectrum availability, and broadcasts the decision back to the \su s through a control channel.
This could typically be done via either {\em hard} or {\em soft} decision rules. The most common soft decision rule is aggregation, where \fc~collects the \rss~values from the \su s and compares their average to a predefined threshold, \ta, to decide on the channel availability.

In hard decision rules, e.g. voting, \fc~combines votes instead of \rss~values. Here, each \su~compares its \rss~value with \ta, makes a local decision (available or not), and then sends to the \fc~its one-bit local decision/vote instead of sending its \rss~value. \fc~applies then a voting rule on the collected votes to make a channel availability decision.
However, for security reasons to be discussed shortly, it may not be desirable to share \ta~with \su s. In this case, \fc~can instead collect the \rss~values from the \su s, make a vote for each \su~separately, and then combine all votes to decide about the availability of the channel.

In this work, we opted for the voting-based decision rule, with \ta~is not to be shared with the \su s, over the aggregation-based rule.
Two reasons for why choosing voting over aggregation: One, aggregation methods are more prone to sensing errors; for example, receiving some erroneous measurements that are far off from the average of the \rss~values can skew the computed \rss~average, thus leading to wrong decision. Two, voting does not expose users to the DLP attack~\cite{li2012location} (which was identified earlier in Section~\ref{sec:related}).
We chose not to share \ta~with the \su s because doing so limits the action scope of malicious users that may want to report falsified \rss~values for malicious and/or selfish purposes.

In this paper, in addition to this 2-party (i.e., \fc~and \su s) cooperative sensing architecture that we just described above, we investigate a 3-party cooperative sensing architecture, where a third entity, called gateway (\gw), is incorporated along with the \fc~and \su s to cooperate with them in performing the sensing task. As we show later, this gateway allows to achieve higher privacy and lesser computational overhead, but of course at its cost.

\subsection{Security Threat Models and Objectives}
We make the following security assumptions:

\begin{assumption} \label{asm:asm1}
(i) \fc~may modify the value of \ta~in different sensing periods to extract information about the \rss~values of \su s; (ii) \gw~executes the protocol honestly but shows interest in learning information about the other parties; (iii) \fc~does not collude with \su s; and (iv) \gw~does not collude with \su s or \fc. 
\end{assumption}

We aim to achieve the following security objectives:
\begin{objective} \label{obj:SecurityObj-1}
(i) Keep \rss~value of each \su~confidential; and (ii) Keep \ta~confidential. This should hold during all sensing periods and for any network membership change.
\end{objective}

\section{Preliminaries}
\label{sec:Preliminaries}

\subsection{Half-Voting Availability Decision Rule}
\label{subsec:half-voting}
Our proposed schemes use the {\em half-voting decision rule}, shown to be optimal in~\cite{zhang2008cooperative}, and for completeness, we here highlight its main idea.
Details can be found in~\cite{zhang2008cooperative}.

Let $\loHyp_0$ and $\loHyp_1$ be the spectrum sensing hypothesis that \pu~is absent and present, respectively. Let $P_f$, $P_d$ and $P_m$ denote the probabilities of false alarm, detection, and missed detection, respectively, of one \su; i.e., $P_f = Pr(\rss>\ta\mid \loHyp_0)$, $P_d = Pr(\rss>\ta\mid \loHyp_1)$, and $P_m = 1 - P_{d}$.

\fc~collects the 1-bit decision $\loDec_i$ from each $\su_i$ and fuses them together according to the following fusion rule~\cite{zhang2008cooperative}:
\begin{equation} \label{FCHyp}
\decision =  \begin{cases}  \Hyp_1, & \displaystyle\sum \limits _{i=1}^n \loDec_i\geq \lam \\  \Hyp_0, & \displaystyle\sum \limits _{i=1}^n \loDec_i <\lam\end{cases}
\end{equation}
\fc~infers that \pu~is present, i.e. $\Hyp_1$, when at least \lam~\su s are inferring $\loHyp_1$. Otherwise, \fc~decides that \pu~is absent, i.e. $\Hyp_0$. Note that the {\em OR} fusion rule, in which \fc~decides $\Hyp_1$ if at least one of the decisions from the \su s is $\loHyp_1$, corresponds to the case where $\lam=1$. The {\em AND} fusion rule, in which \fc~decides $\Hyp_1$ if and only if all decisions from the \su s are $\loHyp_1$, corresponds to the case where $\lam=\nbr$. The cooperative spectrum sensing false alarm probability, $Q_f$, and missed detection probability, $Q_m$, are: $Q_f = Pr(\Hyp_1\mid\loHyp_0)$ and $Q_m = Pr(\Hyp_0\mid\loHyp_1)$.

Letting $n$ be the number of \su s, the optimal value of \lam~that minimizes $Q_f+Q_m$ is $\lam_{opt} = \min(\nbr, \lceil{\nbr}/{(1+\alhv)}\rceil)$,
%
where $\alhv = \ln (\frac{P_f}{1-P_m})/\ln (\frac{P_m}{1-P_f})$ and $\lceil\cdot\rceil$ denotes the ceiling function. 
The value of $\lam_{opt}$ comes from the half-voting rule presented in~\cite{zhang2008cooperative}. We use it since it was proven in~\cite{zhang2008cooperative} to provide the best sensing performance in voting based cooperative sensing.
For simplicity, $\lam_{opt}$ is denoted as $\lam$ throughout this paper.

\subsection{Reputation Mechanism}

To make the voting rule more reliable, we incorporate a reputation mechanism that allows \fc~to progressively eliminate faulty and malicious \su s. It does so by updating and maintaining a reputation score for each \su~that reflects its level of reliability. Our proposed schemes incorporate the {\em Beta Reputation} mechanism~\cite{arshad2011robust}. For completeness, we highlight its key features next; more details can be found in~\cite{arshad2011robust} from which all computations in this subsection are based.

At the end of each sensing period $t$, \fc~obtains a decision vector, $\mbox{\boldmath$ \bin$}(t)=[\bin_1(t),\bin_2(t),\ldots,\bin_\nbr(t)]^T$ with $\bin_i(t) \in \{0,1\}$, where $\bin_i(t)=0$ (resp. $\bin_i(t)=1$) means that the spectrum is reported to be free (resp. busy) by user $\su_i$. \fc~then makes a global decision using the fusion rule $f$ as follows:
\begin{equation}
\label{fDef}
\decision(t)=f(\boldsymbol{w}(t),\mbox{\boldmath$ \bin$}(t)) = \begin{cases} 1 & \mbox{if } \sum\limits_{i=1}^{\nbr}w_i(t) \bin_i(t) \geq\lam \\ 0 & \mbox{otherwise }\end{cases}
\end{equation}
where $\boldsymbol{w}(t) = [w_1(t),w_2(t)\ldots,w_\nbr(t)]^T$ is the weight vector calculated by \fc~based on the credibility score of each user, as will be shown shortly, and $\lam$ is the voting threshold determined by the Half-voting rule~\cite{zhang2008cooperative}, as presented in Section~\ref{subsec:half-voting}.

For each $\su_i$, \fc~maintains positive and negative rating coefficients, $\posr_i(t)$ and $\eta_i(t)$, that are updated every sensing period $t$ as: $\posr_i(t) = \posr_i(t - 1) +\nu_1(t)$ and $\eta_i(t) = \eta_i(t - 1) +\nu_2(t)$, where
$\nu_1(t)$ and $\nu_2(t)$ are calculated as
\vspace{-20pt}
\begin{multicols}{2}
\begin{equation*}
\nu_1(t) = \begin{cases} 1 & \bin_i(t) = \decision(t)\\ 0 & \mbox{otherwise }\end{cases}
\end{equation*}\break
\begin{equation*}
\nu_2(t) = \begin{cases} 1 & \bin_i(t) \neq \decision(t)\\ 0 & \mbox{otherwise }\end{cases}
\end{equation*}
\end{multicols}
Here, $\posr_i(t)$ (resp. $\eta_i(t)$) reflects the number of times $\su_i$'s observation, $\bin_i(t)$, agrees (resp. disagrees) with the $\fc$'s global decision, $\decision$(t).

\fc~computes then $\su_i$'s credibility score, $\varphi_i$(t), and contribution weight, $w_i$(t), at sensing period $t$ as suggested in~\cite{arshad2011robust}:
\vspace{-30pt}
\begin{multicols}{2}
\begin{equation}
\label{cred}
\varphi_i (t) \!=\! \dfrac{\posr_i(t) + 1}{\posr_i(t) \! + \! \eta_i(t)\! +\! 2}
\end{equation}\break
\begin{equation}
\label{weight}
w_i (t)\!= \! {\varphi_i(t)}/{\sum\limits_{j=1}^{\nbr}\!\varphi_j(t)}
\end{equation}
\end{multicols}

\subsection{Cryptographic Building Blocks}
\label{subsec:crypto}
Our schemes use a few known cryptographic building blocks, which we define next before using them in the next sections when describing our schemes so as to ease the presentation.



\begin{definition}~\label{def:OPE}
\noindent \textbf{Order Preserving Encryption~\mbox{\boldmath$(\ope)$}:} is a deterministic symmetric encryption scheme whose encryption preserves the numerical ordering of the plaintexts, i.e. for any two messages $m_1$ and $m_2 \:~s.t.~\: m_1 \leq m_2$, we have $c_1\as\OEnc{K}{m_1}$ $\leq c_2\as\OEnc{K}{m_2}$ \cite{boldyreva2009order}, with $c\as\OEnc{K}{m}$ is order preserving encryption of a message $m\in\{0,1\}^{d}$ under key $K$, where $d$ is the block size of \ope.
\end{definition}

\begin{definition} \label{def:YM}
\noindent \textbf{\em Yao's Millionaires'~\mbox{\boldmath$(\ym)$ Protocol~\cite{yao1982protocols}:}} is a Secure Comparison protocol that enables two parties to execute "the greater-than" function, $GT(x, y) = [x > y]$, without disclosing any other information apart from the outcome.
\end{definition}

\begin{definition} \label{def:GroupKey}
\textbf{\em Tree-based Group Elliptic Curve Diffie-Hellman $\boldsymbol{(\TGECDH)}$~\cite{wang2006performance}:} is a dynamic and contributory group key establishment protocol that permits multiple users to collaboratively establish and update a group key $K$.
\end{definition}
\begin{definition} \label{def:KeyIndependence}
\textbf{Group Key independence:} given a subset of previous keys, an attacker cannot know any other group key.
\end{definition}

\begin{definition} \label{def:ECDLP}
\textbf{\em Elliptic Curve Discrete Logarithm Problem $\boldsymbol{(\ECDLP):}$} given an elliptic curve $\mathscr{E}$ over
$GF(q)$ and points $(P,Z) \in E$, find an integer $x$, if any exists, s.t. $Z = xP$.
\end{definition}
\begin{definition} \label{def:DigSign}
\textbf{\em Digital Signature:} A digital signature scheme \SGN~is used to validate the authenticity and integrity of a message $m$. It contains three components defined as follows: 

$\bullet$ Key generation algorithm ($\mathsf{Kg}$): returns a private/public key pair given a security parameter $1^\kap$, $(\skhors,\pkhors)\leftarrow \mathsf{\SGN.Kg}(1^\kap)$.

$\bullet$ Signing algorithm ($\mathsf{Sign}$): takes as input a message $m$ and the secret key $\skhors$ and returns a signature $\sign$, $\sign \leftarrow \mathsf{\SGN.Sign}(\skhors,m)$.

$\bullet$ Verification algorithm ($\mathsf{Ver}$): takes as input the public key $\pkhors$, $m$ and $\sign$. It returns $1$ if valid and $0$ if invalid, $\{0,1\}\leftarrow \mathsf{\SGN.Ver}(\pkhors,m,\sign)$.
\end{definition}

Note that communications are made over a secure (authenticated) channel maintained with a symmetric key (e.g., via SSL/TLS) to ensure confidentiality and authentication. For the sake of brevity, we will only write encryptions but not the authentication tags (e.g., Message Authentication Codes~\cite{JonathanKatzModernCrytoBook}) for the rest of the paper.

 In the following we present the two schemes that we propose in this paper. For convenience and before getting into the details of the proposed approaches, we have summarized the different notations that we use in the remaining parts of this paper in Table~\ref{t:notations}.
 
 \begin{table}[h!]
\caption{Notations}
\centering
\scriptsize
\resizebox{0.5\textwidth}{!}{
\label{t:notations}
\begin{tabular}{cl}
\hline
\noalign{\medskip }
$\su$ & Secondary user \\
$\fc$ & Fusion center\\
$\gw$ & Gateway\\
$\rss$ & Received signal strength\\
$\gam$ & $=|\rss|=|\ta|$ \\
$\nbr$ & Average number of \su s per sensing period \\
$\mathcal{G}$ & Set of all \su s in the system\\
$\lam$ & Optimal voting threshold \\
$\ta$ & Energy sensing threshold \\
$q$ & Large prime number for \ECELG\\
$\mathscr{E}$ & Elliptic curve over a finite field $GF(q)$ \\
$b_i$ & Outcome of \yme~between $\ta$~and $\rss_i$ \\
$\decision$ & Final decision made by \fc \\
$K$ & Group key established by \su s\\
$\sign$ & Digital signature\\
$\boldsymbol{w}$ & Vector of weights assigned to \su s \\
$T$ & Table of \ECELG~ciphertexts exchanged in \yme \\
$\pkhors$ &  Public key used for the digital signature\\
$\skhors$ &  Secret key used for the digital signature\\
$k_{\fc,i}$ & Secret key established between $\fc~\&~\su_i$ \\
$k_{\gw,i}$ & Secret key established between $\gw~\&~\su_i$ \\
$k_{\fc,\gw}$ & Secret key established between $\fc~\&~\gw$ \\
$(E,D)$ & \ECELG~encryption-decryption for \yme \\
$(\mathcal{E},\mathcal{D})$ & IND-CPA secure block cipher encryption-decryption\\
$\OEncc$ & \ope~encryption \\
$c_i$ & $= \OEnc{K}{\rss_i}$ \\
$\ciphFC_{i}$ & $=\Enc{\key_{\fc,\gw}}{\OEnc{\key_{\fc,i}}{\ta}}$ \\
$\ciphU_{i}$ & $=\Enc{\key_{\gw,i}}{\OEnc{\key_{\fc,i}}{\rss_i}}$ \\
$\ciphGW$ &$ =\Enc{\key_{\fc,\gw}}{\{\bin_i\}_{i=1}^{n}}$ \\
$\chn_i$ & Secure authenticated channel between \fc~and $\su_i$\\
$\LU $ & History list including all values learned by $\{\su_i\}_{i=1}^{n}$\\
$\LF$ & History list including all values learned by \fc\\
$\LG$ & History list including all values learned by \gw\\
$\chr(t)$ & Average number of \su s joining the \crn~at $t$ \\
$\sr$ & Average of the membership change process\\
\noalign{\smallskip} \hline \noalign{\smallskip}
\end{tabular}
}
\end{table}


\section{\ROLPOS}
\label{sec:replpos}

We now present our first proposed scheme, which is a voting-based approach designed for the 2-party cooperative spectrum sensing network, consisting of one \fc~and a set of \su s. Throughout, we refer to this scheme by \ROLPOS~(location privacy for 2-party spectrum sensing architecture).
\ROLPOS~achieves the aforementioned security objectives via an innovative integration of the \ope, \TGECDH~and \ym~protocols.  Voting-based spectrum sensing offers several advantages over its aggregation-based counterparts as discussed in Section \ref{sec:Preliminaries}, but requires comparing \fc's threshold $\ta$ and \su s' \rss s, thereby forcing at least one of the parties to expose its information to the other.
One solution is to use a secure comparison protocol, such as \ym, between \fc~and each \su, which permits \fc~to learn the total number of \su s above/below $\ta$ but nothing else. However, secure comparison protocols involve several costly public key crypto operations (e.g., modular exponentiation), and therefore $\mathcal{O}(\nbr)$ invocations of such a protocol per sensing period, thus incurring prohibitive computational and communication overhead.

$\bullet$ {\em Intuition}: The key observation that led us to overcome this challenge is the following: If we enable \fc~to learn the relative order of \rss~values but nothing else, then the number of \ym~invocations can be reduced drastically. That is, {\em the knowledge of relative order permits \fc~to execute \ym~protocol at worst-case $\mathcal{O}(log(\nbr))$ by utilizing a binary-search type approach}, as opposed to running \ym~with each user in total $\mathcal{O}(\nbr)$ overhead.
This is where \ope~comes into play. {\em The crux of our idea is to make users \ope~encrypt their \rss~values under a group key $K$, which is derived via \TGECDH~at the beginning of the protocol}. This allows \fc~to learn the relative order of encrypted \rss~values but nothing else (and users do not learn each others' \rss~values, as they are sent to \fc~over a pairwise secure channel). \fc~then uses this knowledge to run \ym~protocol by utilizing a {\em binary-search} strategy, which enables it to identify the total number of users above/below $\ta$ and then compares it to $\lam$. As \fc~may try to maliciously modify the value of \ta~as stated in Security Assumption~\ref{asm:asm1}, this makes it easier for it to infer the \rss~values of \su s, thus their location. We rely on digital signatures to overcome this limitation. A digital signature is used by \su s to verify the integrity of the information that was sent by \fc~during the execution of \ym~protocol and signed by the service operator as we explain in more details next. This strategy makes \ROLPOS~achieve \su s' location privacy with efficient spectrum sensing, fault-tolerance and network dynamism simultaneously.

Before we describe our protocol in more details, we first highlight how we improve the \ym~protocol proposed in \cite{lin2005efficient} as shown next.

\subsection{Our Improved \yme~Scheme}
\label{sub:yme}

To achieve high efficiency, we improve the \ym~protocol in \cite{lin2005efficient}, in which only the initiator of the protocol learns the outcome, and call this improved scheme {\em\yme}. \yme, described next, is used by our proposed \ROLPOS~to perform secure comparisons. Our secure comparison scheme improves \ym~protocol proposed in~\cite{lin2005efficient} in two aspects: (i) We adapt it to work with additive homomorphic encryption (specifically \ECELG) to enable compact comparison operations in Elliptic Curves (EC) domain. (ii) The final stage of \yme~requires solving \ECDLP~(Definition \ref{def:ECDLP}), which is only possible with small plaintext domains, and this is the case for our 8-bit encoded RSS values required by {\em IEEE 802.22 standard}~\cite{IEEEStd80222a}. However, despite small plaintext domain, solving \ECDLP~with brute-force is still costly. We improve this step by adapting \POLLAM~method~\cite{blake1999elliptic} to solve the \ECDLP~for the reverse map, which offers decryption efficiency and compactness. The \POLLAM~method is designed to solve the \ECDLP~for points that are known to lie in a small interval, which is the case for \rss~values~\cite{blake1999elliptic}.  Below, we outline our optimized \yme.

$\bullet$ {\em Notation}: Let $\gam=|\rss|=|\ta|$ denote the size in bits of the \rss~value of a \su~and \ta~of \fc~to be privately compared. Also, let $\nbr$ denote the average number of \su s per sensing period, $q$ be a large prime number, $\mathscr{E}$ an elliptic curve over a finite field $GF(q)$, $Z$ a point on the curve with prime order $m$. $(\sk,\pk)$ is a private/public key pair of Elliptic Curve ElGamal (\ECELG) encryption~\cite{koblitz1987elliptic}, generated under $(\mathscr{E},q,Z,m)$. Let $\pr=(\gam,\mathscr{E},q,Z,m,\langle\sk,\pk\rangle)$ be \yme~parameters generated by \fc~which is the initiator of the protocol. \yme~returns $b\as\ymg(\ta,\rss,\pr)$, where $b=0$ if $\ta<\rss$ and $b=1$ otherwise. Only \fc~learns $b$ but $(\fc,\su)$ learn nothing else. For simplicity during the description of \yme, we denote \ta~as $x$ and \rss~as $y$.

\yme, as in \ym, is based on the fact that {\em $x$ is greater than $y$ $\mathit{iff}$ $S^1_x$ and $S^0_y$ have a common element} where $S^1_x$ and $S^0_y$ are the 1-encoding of $x$ and the 0-encoding of $y$ respectively. The 0-encoding of a binary string $s = s_\gam s_{\gam-1}\ldots s_1 \in \{0,1\}^{\gam}$ is given by $S^0_s = \{s_\gam s_{\gam-1}\ldots s_{i+1}1|s_i=0, 1\leq i\leq \gam\}$ and the 1-encoding of $s$ is given by $S^1_s = \{s_\gam s_{\gam-1}\ldots s_i|s_i=1, 1\leq i\leq \gam\}$. For example, if we have a string $s=101101$, then $S^0_s = \{11,10111\}$ and $S^1_s = \{1,101,1011,101101\}$. If we want to compare two values $x=46=101110$ and $y=45=101101$, we need first to construct $S^1_x = \{1,101,1011,10111\}$ and $S^0_y = \{11,10111\}$. Since $S^1_x \cap S^0_y \neq \emptyset$, then $x>y$.

\begin{figure}[h!]
\center
    \includegraphics[width=0.45\textwidth]{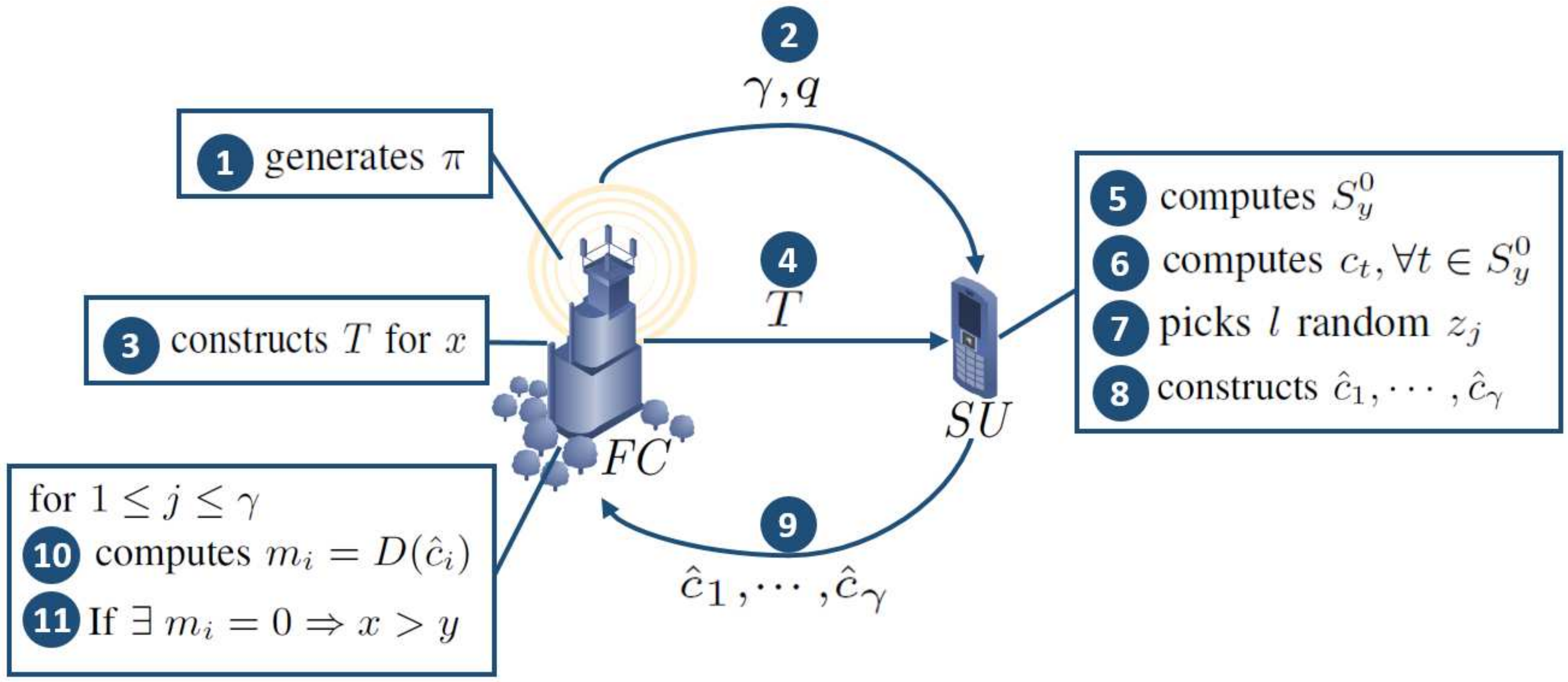}
    \caption{\yme~protocol}
    \label{fig:yme}
\end{figure}

\fc~with a private input $x = x_\gam x_{\gam-1} \ldots x_1$ generates \pr~for encryption and decryption $(E,D)$ then prepares a $2\times \gam$-table $T[i,j]$, $i \in {0,1}, 1\leq j\leq \gam$ such that $T[x_i,i]=E(1)$ and $T[\bar{x_i},i]=E(r_i)$ for a random $r_i$ in the subgroup $G_q$ and finally sends $T$ to \su. \su~with private input $y = y_\gam y_{\gam-1} \ldots y_1$ computes $c_t$ for each $t=t_l t_{\gam-1} \ldots t_i \in S^0_y$ as follows
\begin{equation}
\label{eq:ym}
c_t=T[t_\gam,\gam]\oplus T[t_{\gam-1},\gam-1]\ldots \oplus T[t_i,i]
\end{equation}
with $\oplus$ denotes Elliptic Curve point addition operations ($\oplus$ replaces $\times$ in the original \ym~scheme). \su~then prepares $l=\gam-|S^0_y|$ random encryptions $z_j = (a_j,\bin_j)\in G^2_q, 1\leq j\leq l$ and permutes $c_t$'s and $z_j$'s to obtain $\hat{c}_1,\cdots,\hat{c}_\gam$ which are sent back to \fc~that decrypts $D(\hat{c}_i) = m_i$, $1\leq j\leq \gam$ via \POLLAM~algorithm \cite{blake1999elliptic} and decides $x>y$ $\mathit{iff}$ some $m_i=0$ ($m_i=1$ in the original \ym). The different steps of this protocol are summarized in Figure~\ref{fig:yme}.

\subsection{\ROLPOS~Descitpion}
Next we describe our proposed scheme \ROLPOS~whose main steps are outlined in Algorithm~\ref{alg1}.

\begin{algorithm}[h!]
\caption{\ROLPOS~Algorithm}\label{alg1}
\begin{algorithmic}[1]

\Statex   \textbf{Initialization}: Executed only once.
\State Service operator sets \ta.
\State \fc~generates \pr, sets \lam~and $\boldsymbol{w} \gets \boldsymbol{1}$.
\State \fc~pre-computes $T$ using \pk.
\State Service operator computes $\sign \leftarrow \mathsf{\SGN.Sign}(\skhors,T)$.
\State Service operator shares \pkhors~with \su s.
\State $\mathcal{G}=\{\su_i\}_{i=1}^{\nbr}$ establish $K$ via \TGECDH~protocol.
\State \fc~establishes $\chn_i$~with each $\su_i$ for $i = 1,\ldots,\nbr$.
\hspace{20pt}\algrule
\Statex   \textbf{Private Sensing}: Executed every sensing period $t_w$
\State $\su_i$ computes $c_i \as \OEnc{K}{\rss_i}$ for $i=1,\ldots,\nbr$.
\State $\su_i$ sends $c_i$ to \fc~over $\chn_i$ for $i=1,\ldots,\nbr$.
\State \fc~sorts encrypted RSS values as $c_{min} \leq \ldots \leq c_{max}$.\label{alg1:line:pt}
\State \fc~runs $b_{id_{max}}\as\yme$ $(\rss_{max},\ta,\pr)$ with $\su_{id_{max}}$ having $c_{max}$.\label{alg1:umax}
\State $\su_{id_{max}}$ verifies $T$ using \sign. \label{alg1:verif1}
\If {$\mathsf{\SGN.Ver}(\pkhors,T,\sign)=0$} 
\State $\su_{id_{max}}$ leaves the sensing
\State Go to Step~\ref{alg1:umin}.
\EndIf

\If {$b_{id_{max}}=0$} $\decision$ $ \gets $ Channel free, $\{b_i\}_{i=1}^\nbr \as \boldsymbol{0}$.\label{alg1:decVec1}
\Else \: \fc~runs $b_{id_{min}}\as\yme(\rss_{min},\ta,\pr)$ with $\su_{id_{min}}$ having $c_{min}$. \label{alg1:umin}
\State $\su_{id_{min}}$ verifies $T$ using \sign. \label{alg1:verif2}
\If {$\mathsf{\SGN.Ver}(\pkhors,T,\sign)=0$} 
\State $\su_{id_{min}}$ leaves the sensing
\State Go to Step~\ref{alg1:all}.
\EndIf
\If {$b_{id_{min}}=1$}  $\decision$ $ \gets $ Channel busy, $\{b_i\}_{i=1}^\nbr \as \boldsymbol{1}$.\label{alg1:decVec2}\Else \Repeat  
 \State \fc~computes $I \gets BinarySearch(\mathcal{G})$
 \State \fc~runs $b_I\as\yme$ $(\rss_I,\ta,\pr)$ with $\su_I$ having $c_{I}$.\label{alg1:ui}
 \State $\su_I$ verifies $T$ using \sign.\label{alg1:all} 
\If {$\mathsf{\SGN.Ver}(\pkhors,T,\sign)=0$} 
\State $\su_I$ leaves the sensing
\EndIf
 
\Until {$\rss_{I-1} \leq \ta \leq \rss_I$}

\State \fc~assigns $b_i\as 0$\:for\:$i=1,\ldots,I-1$\:and\: $b_j\as 1$ for $j=I,\ldots,\nbr$ \label{alg1:decVec3}
\State \fc~computes $\votes\as \sum\limits_{i=1}^{\nbr}w_i\times b_i$ \label{alg1:rep1}
\If {$\votes \geq  \lam$} $\decision$ $ \gets $ Channel busy
\Else \: $\decision$ $ \gets $ Channel free
\EndIf
\EndIf
\EndIf
\State \fc~updates $\{\varphi_i\}_{i=1}^\nbr$ and $\{w_i\}_{i=1}^\nbr$ as in Eqs.~\eqref{cred} \& \eqref{weight}\label{alg1:rep3}

\Return $\decision$
\hspace{20pt}\algrule
\Statex \textbf{Update after $\mathcal{G}$ Membership Changes or Breakdown}: \If {\su(s) join/leave $\mathcal{G}$ or breakdown in $t_w$} 
\State New group  $\mathcal{G}'$ form new $K'$ using \TGECDH. 
\State \fc~updates \lam~and \pr~as \lam'~and \pr',~respectively, if required. 
\State Execute the private sensing with $(K',\lam',\pr')$.
\EndIf

\end{algorithmic}
\end{algorithm}

 $\bullet$~{\em Initialization:}  The service operator sets up the value of energy threshold \ta. \fc~sets up \ECELG~crypto parameters, voting threshold and users reputation weights values. Initially, all the users are considered credible so the weight vector $\boldsymbol{w}$ is constituted of ones. \fc, then, constructs the table $T$ used in \ym~protocol as described in Section~\ref{sub:yme} with \ta~as input using the \fc's \ECELG~public key \pk. Notice here that since the same \ta~is always used during different sensing periods, the table $T$ can be precomputed during the {\em Initialization} phase. This considerably reduces this protocol's computational overhead. Then the service operator that manages the network signs $T$ using a digital signature scheme with secret key \skhors. This digital signature is used to make sure that \fc~does not maliciously modify the value of \ta~to learn \rss~values of users and thus infer their locations. The service operator then shares the public key \pkhors~with \su s to use it for verifying the integrity of $T$ and thus of \ta. \su s establish a group key $K$ via  \TGECDH, with which they \ope~encrypt their \rss~values during the private sensing. \fc~also establishes a secure channel $\chn_i$ with each user $\su_i$.

 $\bullet$~{\em Private Sensing:} Each $\su_i$ \ope~encrypts its $\rss_i$ with group key $K$ and sends ciphertext $c_i$ to \fc~over $\chn_i$. \fc~then sorts ciphertexts as $c_{min} \leq \ldots \leq c_{max}$ (as all $\rss_i$s are \ope~encrypted under the same $K$) without learning corresponding \rss~values, and the secure channel $\chn_i$ protects the communication of $\su_i$ from other users as well as from outside attackers. \fc~then initiates \yme~first with the $\su_{id_{max}  }$~that has the highest \rss~value $\rss_{max}$. If it is smaller than energy sensing threshold $\ta$, then the channel is free. Otherwise, \fc~initiates \yme~with the user that has $\rss_{min}$. If it is greater than $\ta$, then the channel is busy. Otherwise, to make the final decision based on the optimal sensing threshold $\lam$, \fc~runs \yme~according to the binary-search strategy which guarantees the decision at the worst $\mathcal{O}(log(\nbr))$ invocations. Note that before participating in \yme, each \su~first verifies the integrity of $T$ using the digital signature \sign~that was provided by the service operator as indicated in Steps~\ref{alg1:verif1}, \ref{alg1:verif2} $\&$ \ref{alg1:all}. A \su~that detects a change in the value of $T$ refuses to participate in the sensing to prevent \fc~from learning any sensitive information regarding its location. In that case the system stops and the malicious intents of \fc~are detected.
 
\begin{figure}[h!]
\center
    \includegraphics[width=0.46\textwidth,height = 9.2cm]{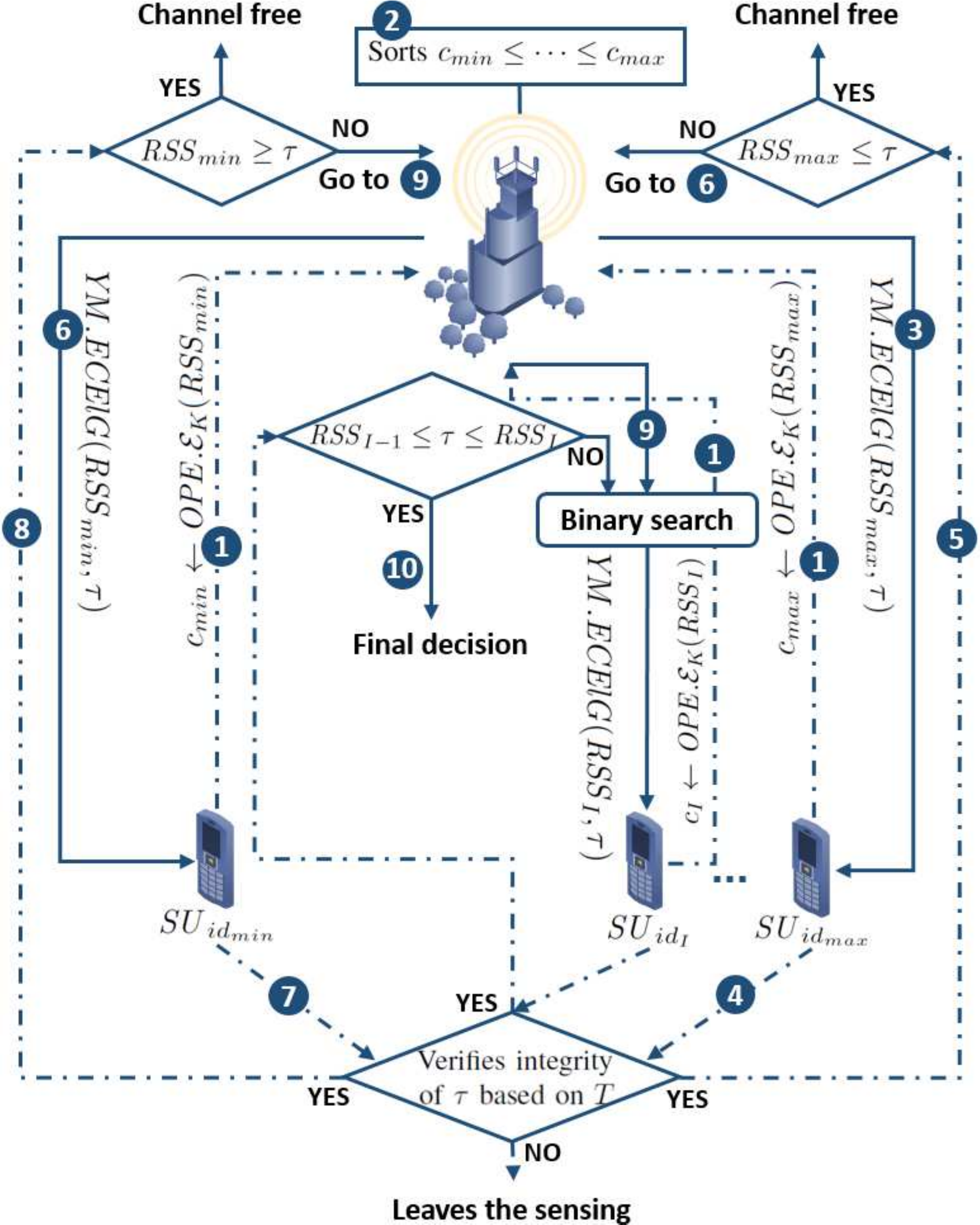}
    \caption{\ROLPOS's Private Sensing phase}
    \label{Fig:LP-2PSS}
\end{figure}

 In Steps~\ref{alg1:decVec1}, \ref{alg1:decVec2} $\&$ \ref{alg1:decVec3} of Algorithm~\ref{alg1}, \fc~constructs the vector of local decisions of \su s after running the private comparisons between \ta~and \rss~values. Based on the decision vector $\boldsymbol{\bin}$ and the weights vector $\boldsymbol{w}$ that was computed previously, \fc~computes $\votes$ in Step \ref{alg1:rep1} using Equation~\ref{fDef} to finally make the final decision \decision~using voting threshold $\lam$. \fc~then computes the credibility score and the weights that will be given to all users in the next sensing period. If $\su_i$ has a decision $b_i \neq \decision$, its assigned weight decreases. But if a \su~makes the same decision as \fc, it is assigned the highest weight. The main steps of the private sensing phase are summarized in Figure~\ref{Fig:LP-2PSS}.

$\bullet$~{\em Update after $\mathcal{G}$ Membership Changes or Breakdown:} At the beginning of $t_w$, if membership status of $\mathcal{G}$ changes, a new group key is formed via \TGECDH, and then \fc~updates \lam. If some \su s breakdown and fail to sense or send their measurements, \lam~also must be updated. In new sensing period, Algorithm \ref{alg1} is executed with new parameters and group key.

\subsection*{Choice of digital signature} 
Choosing the right digital signature scheme depends on the network and users constraints. In the following we briefly discuss some of the schemes that could be applied in \ROLPOS.

One scheme that could be used is {\em RSA} \cite{Rivest1978MOD} which is one of the first and most popular digital signature schemes. {\em RSA} has a very large signature but offers a fast signature verification. However, newer schemes outperform it in terms of signature and key size and/or computational efficiency.

Another scheme could be {\em ECDSA} \cite{johnson2001elliptic} which is an elliptic curve analogue of the {\em DSA} \cite{fips1994186} digital signature scheme. It provides more compact signatures than its counterparts thanks to the use of Elliptic Curve crypto. It has a moderate speed,  though, in terms of verification and encryption compared to {\em RSA}. It is more suitable for situations where the communication overhead is the main concern.

One-time signatures, e.g. \HORS~\cite{reyzin2002better} and its variants \cite{neumann2004horse,pieprzyk2004multiple}, are digital signatures that are based on one-way functions without a trapdoor which makes them much faster than commonly used digital signatures, like {\em RSA}. The main drawbacks of this kind of digital signatures are their large size and the complexity of their "one-timed-ness" which requires a new call to the key generation algorithm for each use. In our context, we should not worry about the latter since we sign $T$ only once so we don't have to regenerate the keys. In that case, one-time signatures may be the best option when computation speed at \su s is the main concern.

{\em NTRU} \cite{hoffstein2003ntrusign} signature could also be applied here. It provides a tradeoff between signature size and computational efficiency. Indeed it has a moderate signature size that is larger than the one of {\em ECDSA} but it is faster than both {\em ECDSA} and {\em RSA} in key generation, signing and verification.

\section{\LPGW}
\label{sec:lpgw}
We now present an alternative scheme that we call \LPGW~(location privacy for 3-party spectrum sensing architecture), which offers higher privacy and significantly better performance than that of \ROLPOS, but at the cost of deploying an additional entity in the network, referred to as Gateway (\gw) (thus "3P" refers to the 3 parties: \su s, \fc, and \gw).

 \gw~enables a higher privacy by preventing \fc~from even learning the order of encrypted \rss~values of \su s (as in \ROLPOS). \gw~also learns nothing but secure comparison outcome of a \rss~values and \ta, as in \ym~but only using \ope. Thus, no entity learns any information on \rss~or~\ta~beyond a pairwise secure comparison, which is the minimum information required for a voting-based decision.

%

$\bullet$ {\em Intuition}: The main idea behind \LPGW~is simple yet very powerful: We enable \gw~to privately compare \nbr~distinct \ope~encryptions of \ta~and \rss~values, which were computed under \nbr~pairwise keys established between \fc~and \su s. These \ope~encrypted pairs permit \gw~to learn the comparison outcomes without deducing any other information. \gw~then sends these comparison results to \fc~to make the final decision. \fc~learns no information on \rss~values and \su s cannot obtain the value of \ta, which complies with our Security Objectives \ref{obj:SecurityObj-1}. Note that \LPGW~relies {\em only on symmetric cryptography} to guarantee the location privacy of \su s. Hence, it is the {\em most computationally efficient and compact} scheme among all alternatives but with an additional entity in the system. \LPGW~is described in Algorithm~\ref{alg2} and  outlined below. 

 \begin{algorithm}[h!]
\caption{\LPGW~Algorithm}\label{alg2}
\begin{algorithmic}[1]
\Statex  \textbf{Initialization}: Executed only once.
\State Service operator sets \ta.
\State \fc~sets \lam~and $\boldsymbol{w} \gets \boldsymbol{1}$.

\State  \fc~establishes $\key_{\fc,i}$ with $\su_i$, $i=1,\ldots ,\nbr$.
\State  \gw~establishes $\key_{\gw,i}$ with $\su_i$, $i=1,\ldots ,\nbr$.
\State  \fc~establishes $\key_{\fc,\gw}$ with \gw.
\State \fc~computes $\ciphFC_{i} \gets \Enc{\key_{\fc,\gw}}{\OEnc{\key_{\fc,i}}{\ta}}$, $i=1,\ldots ,\nbr$ and sends $\{\ciphFC_i\}_{i=1}^{n}$ to \gw. \label{alg1:enc1}
\hspace{20pt}\algrule
\Statex   \textbf{Private Sensing}: Executed every sensing period $t_w$
\State $\su_i$ computes $\ciphU_{i} \gets \Enc{\key_{\gw,i}}{\OEnc{\key_{\fc,i}}{\rss_i}}$, $i = 1,\ldots,\nbr$ and sends $\{\ciphU_i\}_{i=1}^{n}$ to \gw.\label{alg1:enc2}
\State \gw~obtains $\OEnc{\key_{\fc,i}}{\ta} \gets \Dec{\key_{\fc,\gw}}{\ciphFC_{i}}$ and   $\OEnc{\key_{\fc,i}}{\rss_i} \gets \Dec{\key_{\gw,i}}{\ciphU_{i}}$, $i = 1,\ldots,\nbr$.
\For{\texttt{$i = 1,\ldots,\nbr$}}
\If {$\OEnc{\key_{\fc,i}}{\rss_i} < \OEnc{\key_{\fc,i}}{\ta}$} $\bin_i \gets 0$\label{alg1:comp}
\Else \:$\bin_i \gets 1$
\EndIf
\EndFor
\State \gw~computes $\ciphGW \gets \Enc{\key_{\fc,\gw}}{\{\bin_i\}_{i=1}^{n}}$ and sends $\ciphGW$ to \fc. \label{alg1:enc}
\State \fc~decrypts $\ciphGW$ and computes $\votes\as \sum\limits_{i=1}^{\nbr}w_i\times b_i$ \label{alg1:sum}

\If {$\votes \geq  \lam$}  $\decision$ $ \gets $ Channel busy
\Else \: $\decision$ $ \gets $ Channel free
\EndIf
\State \fc~updates $\{\varphi_i\}_{i=1}^\nbr$ and $\{w_i\}_{i=1}^\nbr$ as in Eqs.~\eqref{cred} \& \eqref{weight}\label{upWeight}

\Return $\decision$
\hspace{20pt}\algrule
\Statex   \textbf{Update after $\mathcal{G}$ Membership Changes or Breakdown}:
\If{$\su_j$ joins \crn}
\State $\su_j$ establishes $\key_{\fc,j}$ with \fc~and $\key_{\gw,j}$ with \gw.
\EndIf

\If{\su s join/leave/breakdown}
\State \fc~updates \lam~as \lam'.
\State Execute the private sensing with \lam'.
\EndIf

\end{algorithmic}
\end{algorithm}

 $\bullet$~{\em Initialization:} Service operator and \fc~set up spectrum sensing and crypto parameters. Let $(\En,\De)$ be IND-CPA secure~\cite{JonathanKatzModernCrytoBook} block cipher (e.g. \aes) encryption/decryption operations. \fc~establishes a secret key with each \su~and \gw. \gw~establishes a secret key with each \su. \fc~encrypts \ta~with \ope~using $\key_{\fc,i}$, $i=1 \ldots \nbr$. \fc~then encrypts \ope~ciphertexts with \En~using $\key_{\fc,\gw}$ and sends these $\ciphFC_{i}$s to \gw, $i=1 \ldots \nbr$. Since these encryptions are done offline at the beginning of the protocol, they do not impact the online private sensing phase. \fc~may also pre-compute a few extra encrypted values in the case of new users joining the sensing.

 $\bullet$~{\em Private Sensing:} Each $\su_i$~encrypts $\rss_i$ with \ope~using $\key_{\fc,i}$, which was used by \fc~to \ope~encrypt \ta~value. $\su_i$ then encrypts this ciphertext with \En~using key $\key_{\gw,i}$, and sends the final ciphertext $\ciphU_{i}$ to \gw. \gw~decrypts $2\nbr$ ciphertexts $\ciphFC_{i}$s and $\ciphU_{i}$s with \De~using $\key_{\fc,\gw}$ and $\key_{\gw,i}$, which yields \ope~encrypted values. \gw~then compares each \ope~encryption of \rss~with its corresponding \ope~encryption of \ta. Since both were encrypted with the same key, \gw~can compare them and conclude which one is greater as in Step~\ref{alg1:comp}. \gw~stores the outcome of each comparison in a binary vector $\mbox{\boldmath$ \bin$}$, encrpyts and sends it to \fc. Finally, \fc~compares the summation of votes \votes~to the optimal voting threshold \lam~to make the final decision about spectrum availability and updates the reputation scores of the users.

  $\bullet$~{\em Update after $\mathcal{G}$ Membership Changes or Breakdown:} Each new user joining the sensing just establishes a pairwise secret key with \fc~and \gw. This has no impact on existing users. If some users leave the network, \fc~and \gw~remove their secret keys, which also has no impact on existing users. In both cases, and also in the case of a breakdown or failure, \lam~must be updated accordingly. 
  
  \begin{figure}[h!]
\center
    \includegraphics[width=0.45\textwidth]{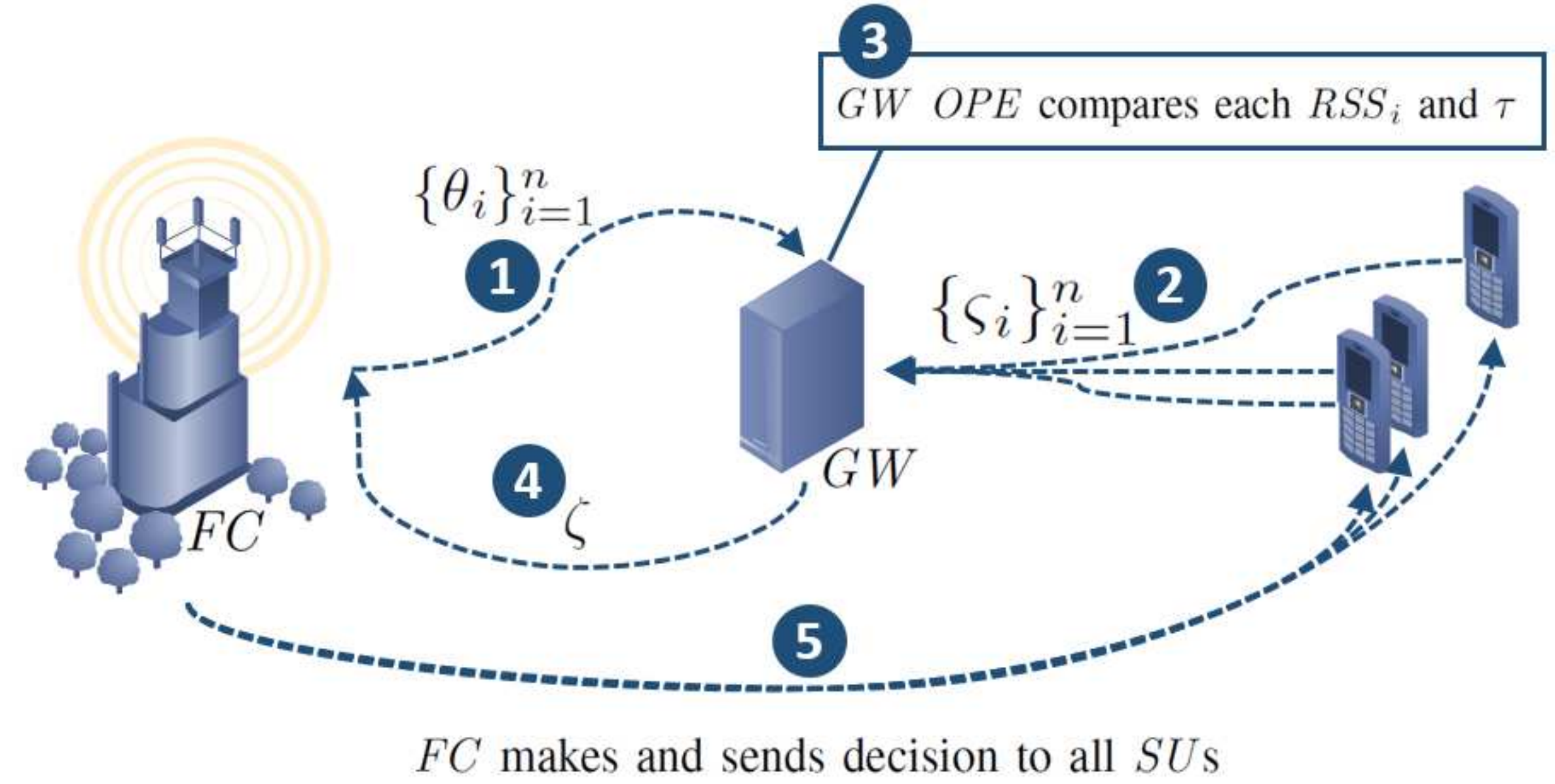}
    \caption{\LPGW~protocol, $\ciphFC_{i} \gets \Enc{\key_{\fc,\gw}}{\OEnc{\key_{\fc,i}}{\ta}}$, $\ciphU_{i} \gets \Enc{\key_{\gw,i}}{\OEnc{\key_{\fc,i}}{\rss_i}}$ and $\ciphGW \gets \Enc{\key_{\fc,\gw}}{\{\bin_i\}_{i=1}^{n}}$}
    \label{yme}
\end{figure}
  \begin{myremark}
A malicious \fc~in \LPGW~following Security Assumption~\ref{asm:asm1} may want to maliciously modify the value of \ta. But since \gw~is the one that performs the comparison between \rss~values and \ta, changing \ta~maliciously has almost no benefit to \fc~as it does not have access to individual comparison outcomes. This makes \LPGW~robust against this malicious \fc.\\
 \end{myremark}
 
 It is worth iterating that the \gw~only needs to perform simple comparison operations between the \rss~values of the \su s and the energy sensing threshold \ta~of the \fc~as we explained earlier. Thus, such an entity does not interfere with the spectrum sensing process in the \crn. Moreover, it does not need to be provided with large computational resources as these comparisons are very simple and fast to perform. It could be a standalone entity, one of the \su s that is dedicated to perform the tasks of the \gw~or even a secure hardware that is deployed inside the \fc~itself as we discuss next. This gives multiple options to system designers. If FCC's regulation allows introducing an additional entity to the \crn, then \gw~could be deployed without any concern. If not, system designers could consider introducing a secure hardware within \fc~or dedicating one of the \su s to perform the tasks of \gw.

  \subsection*{\LPGW~with Secure Hardware}
  \LPGW~could also be implemented in a slightly different way by relying on a {\em secure hardware} deployed within the \fc~itself instead of using a dedicated gateway. All the computation that is performed by \gw~could be relayed to this hardware. This {\em secure hardware}, which is referred to as {\em secure co-processor} (\SCPU) or as {\em trusted platform module} ({\em TPM}) in the literature, is physically shielded from penetration, and the I/O interface to the module is the only way to access the internal state of the module \cite{yee1995secure}. An \SCPU~that meets the FIPS 140-2 level 4 \cite{fips2001140} physical security requirements guarantees that \fc~cannot tamper with its computation. Any attempt to tamper with this \SCPU~from \fc~that results somehow in penetrating
the shield, leads to the automatic erasure of sensitive memory areas containing critical secrets.
  
  The {\em SCPU} may provide several benefits to the network. First, there is no need anymore of adding a new standalone entity managed by a third party to the network as was the case with \gw. Also, despite its high cost, having an \SCPU~deployed within \fc~itself may reduce the communication latency that is incurred by having a gateway that needs to communicate with \fc~and with every user in the network.

In terms of performance, it was proven in \cite{bajaj2014trusteddb} that at a large scale the computation inside an \SCPU~is orders of magnitude cheaper than equivalent cryptography that is performed on an unsecured server hardware, despite the overall greater acquisition cost of secure hardware.

All of this makes using an {\em SCPU} a good alternative to using a dedicated gateway in the network thanks to its performance and the security guarantees that it provides.

\section{Security Analysis}
\label{sec:SecAnalysis}
We first describe the underlying security primitives, on which our schemes rely, and then precisely quantify the information leakage of our schemes, which we prove to achieve our Security Objectives \ref{obj:SecurityObj-1}. At the end of this section, we discuss the security of the modified versions of our schemes.

\begin{fact}\label{fact:IdealSecOPE}
 An \ope~is \emph{indistinguishable under ordered chosen-plaintext attack (IND-OCPA)} \cite{boldyreva2009order} if it has no leakage, except the order of ciphertexts (e.g. \cite{popa2013ideal,kerschbaum2014optimal}).
\end{fact}

\begin{fact}\label{fact:YME}
\yme~is secure by Definition \ref{def:YM} if \ECELG~cryptosystem~\cite{koblitz1987elliptic}, whose security relies on the \ECDLP~(Definition \ref{def:ECDLP}), is secure.
\end{fact}

\begin{fact}\label{fact:SecyreGHD}
\TGECDH~is secure with key independence by Definition \ref{def:KeyIndependence} if \ECDLP~is intractable by Definition \ref{def:ECDLP}.
\end{fact}

Let \En~and \OEncc~be {\em IND-CPA secure}~\cite{JonathanKatzModernCrytoBook} and {\em IND-OCPA secure} symmetric ciphers, respectively. $(\{\rss_{i}^{j}\}_{i=1,j=1}^{n,\ell},\tau)$ are \rss~values and \ta~of each $\su_i$ and \fc~for sensing periods $j=1,\ldots,\ell$ in a group $\mathcal{G}$. $(\LU,\LF,\LG)$ are history lists, which include all values learned by entities $\su_i$, \fc~and \gw, respectively, during the execution of the protocol for all sensing periods and membership status of $\mathcal{G}$. Vector \V~is a list of IND-CPA secure values transmitted over secure (authenticated) channels. \V~may be publicly observed by all entities including external attacker \A. Hence, \V~is a part of all lists $(\LU,\LF,\LG)$. Values (jointly) generated by an entity such as cryptographic keys or variables stored only by the entity itself (e.g., \lam, \pr) are not included in history lists for the sake of brevity. Moreover, information exchanged during the execution of \yme~protocol are not included in history lists, since they do not leak any information by Fact \ref{fact:YME}.

\begin{mytheorem} \label{the:Security:2P}
Under Security Assumptions \ref{asm:asm1}, \ROLPOS~leaks no information on $(\{\rss_{i}^{j}\}_{i=1,j=1}^{n,\ell},\tau)$ beyond IND-CPA secure $\{\V^{j}\}_{j=1}^{\ell}$, IND-OCPA secure order of tuple $(\{\Z^{j}=\OEnc{K^{j}}{\rss_{1}^{j}},\ldots,\OEnc{K^{j}}{\rss_{n}^{j}}\}_{j=1}^{\ell},\ta)$ and $\{b_{i}^{j}\}_{i=1,j=1}^{n,\ell}$ to \fc.
\end{mytheorem}
\noindent {\em Proof:} $\V^{j}=\{\chn_{i}^{j}\}_{i=1,j=1}^{n,\ell}$ at Step 6 of Algorithm \ref{alg1}.  History lists are as follows for each sensing period $j=1,\ldots,\ell$:
\begin{eqnarray*} \label{eq:HistorList3P}
\LU=\V^{j},~~~\LF = (\{b_{i}^{j}\}_{i=1}^{n},\V^{j},\Z^{j}),
\end{eqnarray*}
where $\{b_{i}^{j}\}_{i=1}^{n}$ are the outcomes of \yme~protocol (Steps~\ref{alg1:umax}, \ref{alg1:umin} \& \ref{alg1:ui} of Algorithm \ref{alg1}). By Fact \ref{fact:YME}, \yme~protocol leaks no information beyond $\{b_{i}^{j}\}_{i=1}^{n}$ to \fc~and no information to anyone else. Variables in $(\LU,\LF)$ are IND-CPA and IND-OCPA secure, and therefore leak no information beyond the order of tuples in $\Z^{j}$ to \fc~by Fact~\ref{fact:IdealSecOPE}.

Any membership status update on $\mathcal{G}$ requires an execution of \TGECDH~protocol, which generates a new group key $\Kb^{j}$. By Fact \ref{fact:SecyreGHD}, \TGECDH~guarantees key independence property (Definition \ref{def:KeyIndependence}), and therefore $\Kb^{j}$ is only available to new members and is independent from previous keys.  Hence, history lists $(\LU,\LF)$ are computed identically as described above for the new membership status of $\mathcal{G}$ but with $\Kb^{j}$, which are IND-CPA secure and IND-OCPA secure.\hfill

Using a digital signature gives \su s the possibility to learn the intentions of \fc~and detect whether it is trying to locate them. Since no \su~wants its location to be revealed, \su s will simply refuse to participate in the sensing upon detection of malicious activity of \fc~by verifying the signed messages. The only way that \fc~can learn the location of a \su~in this case is when this \su~continues to participate in the sensing even after detecting the malicious intents of \fc. \hfill$\square$

\begin{mytheorem} \label{the:Security:3P}
Under Security Assumptions \ref{asm:asm1}, \LPGW~leaks no information on $(\{\rss_{i}^{j}\}_{i=1,j=1}^{n,\ell},\tau)$ beyond IND-CPA secure $\{\V^{j}\}_{j=1}^{\ell}$, IND-OCPA secure pairwise order  $\{\OEnc{\key_{\fc,i}}{\rss_{i}^{j}},$ $\OEnc{\key_{\fc,i}}{\ta}\}_{i=1,j=1}^{n,\ell}$ to \gw~and $\{b_{i}^{j}\}_{i=1,j=1}^{n,\ell}$ to \fc.
\end{mytheorem}
\noindent {\em Proof:} $\V^{j}=\{\ciphFC_{i}^{j},\ciphU_{i}^{j},\ciphGW^{j}\}_{i=1,j=1}^{n,\ell}$, where $\{\ciphFC_{i}^{j}\}_{i=1,j=1}^{n,\ell}$ and $\{\ciphU_{i}^{j},\ciphGW^{j}\}_{i=1,j=1}^{n,\ell}$ are generated at the initialization and private sensing in Algorithm~\ref{alg2}, respectively. History lists are as follows for each sensing period $j=1,\ldots,\ell$:
\begin{eqnarray*} \label{eq:HistorList3P}
\LU & = & \V^{j},~~~\LF = (\{b_{i}^{j}\}_{i=1,j=1}^{n,\ell},\V^{j}), \\
\LG & = & (\{\OEnc{\key_{\fc,i}}{\rss_{i}^{j}},\OEnc{\key_{\fc,i}}{\ta}\}_{i=1,j=1}^{n,\ell},\V^{j},\\& &\{b_{i}^{j}\}_{i=1,j=1}^{n,\ell})
\end{eqnarray*}
Variables in $(\LU,\LF,\LG)$ are IND-CPA secure and IND-OCPA secure, and therefore leak no information beyond the pairwise order of ciphertexts to \gw~by Fact \ref{fact:IdealSecOPE}.

Any membership status update on $\mathcal{G}$ requires an authenticated channel establishment or removal for joining or leaving members, whose private keys are independent from each other. Hence, history lists $(\LU,\LF,\LG)$ are computed identically as described above for the new membership status of $\mathcal{G}$, which are IND-CPA secure and IND-OCPA secure.\hfill$\square$

\begin{mycorollary} \label{Cor:SecurityObjectives}
Theorem \ref{the:Security:2P} and Theorem \ref{the:Security:3P} guarantee that in our schemes, RSS values and \ta~are IND-OCPA secure for all sensing periods and membership changes. Hence, our schemes achieve Objectives \ref{obj:SecurityObj-1}.
\end{mycorollary}
\vspace{-5mm}

\subsection{Discussion about {\em SCPU}-based \LPGW's security}
The security of {\em SCPU}-based \LPGW~could be reduced to that of the {\em SCPU} that is used. Since no direct communication exists between \fc~and \su s, the only way for \fc~to learn \rss~values of \su s is by compromising the {\em SCPU}. Having the secret keys that were used to \ope~encrypt \su s' \rss~values, a successful attempt to break into this secure hardware by \fc~will allow it to decrypt the \rss~values and learn \su s' locations. However, as mentioned earlier, a \SCPU~that complies with the physical security requirements of FIPS 140-2 level 4 \cite{fips2001140} should guarantee that such a breach does not happen. And thus, \fc~should not be able to retrieve the data in the {\em SCPU} even though the latter is deployed inside the malicious \fc~itself.

\subsection{Discussion about collusion between different entities}
 We also investigate how our schemes perform under collusion. We discuss different collusion scenarios for each proposed scheme separately.

For \ROLPOS, if multiple \su s collude to learn another \su's location information, their collusion can only allow them to learn IND-CPA secure values \V~which contain the \ope~encrypted \rss s transmitted over the authenticated secure channel between the target \su~and \fc. This means that collusion among \su s does not allow them to learn \rss~measurements of other \su s and, thus, nor their location. The second scenario is when \fc~colludes with some \su s to localize other \su s. In this case, \fc~will have access to the group key $K$ used by \su s to encrypt their \rss~measurements. Only in this case would \fc~be able to learn \su s' locations. Therefore, \ROLPOS~is robust against collusion among compromised \su s, but assumes that \fc~cannot collude with \su s.

Similar reasoning applies to \LPGW. Collusion among \su s does not allow them to infer other \su s' locations. And if \su s collude with \gw, they can only manage to learn the \ope~encrypted \rss~measurements of the \su s but nothing more, as each \su~\ope~encrypts its \rss~measurement with its own secret key. Also, collusion between \fc~and some \su s cannot reveal the \rss~measurements of other \su s as the latter send their \ope~encrypted \rss s through their authenticated channels established individually with the \gw. This prevents colluding \su s and \fc~from accessing the private information of other \su s and subsequently localizing them. Thus, for \LPGW, only collusion between \gw~and \fc~could reveal \rss~measurements of all \su s as \fc~has the secret keys that were used by \su s to \ope~encrypt their \rss~values before sending them to \gw. However, this specific collusion scenario could be dealt with, for example, by deploying a secure hardware within \fc~to play the role of \gw. The inherent nature of such a hardware prevents \fc~from accessing it and colluding with it. We provided an explanation to this in Section~\ref{sec:lpgw}. This proofs that \LPGW~is not only robust against collusion among \su s themselves, but also against collusion between \fc~and compromised \su s.

\begin{table*}[ht!]

\scriptsize
\centering  \caption{Computational overhead comparison} \label{tab:Table2}
\resizebox{\textwidth}{!}{%
\renewcommand{\arraystretch}{1.25}{
\begin{tabular}{||c||c|c|c|c||}

\hline \multicolumn{1}{||c||}{\multirow{2}{*}{\textbf{\em Scheme}}}  & \multicolumn{4}{|c||}{\textbf{Computation}} \\ \cline{2-5}

\multicolumn{1}{||c||}{} &  \textbf{ {\em FC}} & \multicolumn{2}{c|}{\textbf{ {\em SU}}} & \textbf{ {\em GW}}\\ \hline
\hline  \multicolumn{1}{||c||}{\textbf{\ROLPOS}}  &  $\gam/2 \cdot(2+log\:\nbr)\cdot (PMulQ + PAddQ + \sqrt{\delta} \cdot \pol)$ & \multicolumn{2}{c|}{$(4\gam-6)\cdot PAddQ+\ope +\sr\cdot(2\:log\:\nbr +2)\cdot PMulQ$ } & - \\
\hline  \multicolumn{1}{||c||}{\textbf{\LPOS}}  &  $1/2 \cdot(2+log\:\nbr)\cdot\gam \cdot|p|\cdot Mulp$ & \multicolumn{2}{c|}{$(2\gam\cdot |p|+2\gam)\cdot Mulp+\ope +2\sr\cdot log\: \nbr \cdot PMulQ$} & - \\
\hline  \multicolumn{1}{||c||}{\ECEG} &$PMulQ + PAddQ + \sqrt{\nbr\cdot \delta} \cdot \pol$ & $\qquad 2PMulQ +PAddQ \qquad$ & $(\nbr-2)\cdot PAddQ$ & - \\
 \hline \multicolumn{1}{||c||}{\PPSS} &  $H + (\nbr+2) \cdot Mulp + (2^{\gam-1}\cdot\nbr + 2) \cdot Expp$ & \multicolumn{2}{c|}{$H + 2Expp + Mulp$} & -\\ \hline

 \hline \hline   \multicolumn{1}{||c||}{\textbf{\LPGW}}  &  $\De + \chr(t)\cdot(\En+\ope_E)$ & \multicolumn{2}{c|}{$\ope_E + \En$ }& $\nbr \cdot\De + \En$  \\ \hline
\multicolumn{1}{||c||}{\PDAFT} & $2Exp\rsa^2 + Inv\rsa^2 + \srv \cdot Mul\rsa^2$ &\multicolumn{2}{c|}{$ 2Exp\rsa^2 + Mul\rsa^2$ } & $\nbr\cdot Mul\rsa^2$\\ \hline
\end{tabular}}}

\begin{flushleft}
\scriptsize {\doublespacing \textbf{(i) Variables:} $\kap$ security parameter, $\rsa$: modulus in Paillier, $p$: modulus of El Gamal, $H$: cryptographic hash operation, $K$: secret group key of \ope. $Expu$ and $Mulu$ denote a modular exponentiation and a modular multiplication over modulus $u$ respectively, where $u \in \{\rsa, \rsa^2, p\}$. $Inv\rsa^2$: modular inversion over $\rsa^2$, $PMulQ$: point multiplication of order $Q$, $PAddQ$: point addition of order $Q$. $\srv$: number of servers needed for decryption in \PDAFT.
\textbf{(ii)  Parameters size:} For a security parameter $\kappa = 80$, suggested parameter sizes by {\em NIST 2012} are given by : $|\rsa| = 1024$, $|p| = 1024$, $|Q|=192$ as indicated in \cite{keylength}. \textbf{(iii) YM.ECElGamal:} The communication cost for one comparison is $4\gam \cdot |Q|$. The total computational cost of the scheme for one comparison is $\gam\cdot (PMulQ+5PAddQ+\sqrt{\delta} \cdot \pol)-6PAddQ$. \textbf{(iv) ECEG:} The decryption of the aggregated message in \ECEG~is done by solving the constrained ECDLP problem on small plaintext space similarly to \cite{li2012location} via Pollard's Lambda algorithm, which requires $O(\sqrt{n \cdot\delta})\cdot \pol$ computation and $O(log(n\delta))$ storage \cite{menezes2010handbook}, where $\delta = a-b$ if $RSS \in [a,b]$ and \pol \;is the number of point operations in Pollard Lambda algorithm which varies depending on algorithm implementation used. For \su's overhead, the left column shows the cost for a normal \su~in \ECEG~and the right column shows the cost of the \su~that plays the role of a gateway in \ECEG. \textbf{(v) TGECDH}: It permits the alteration of group membership (i.e., join/leave), on average $\mathcal{O}(log(n))$ communication and computation (i.e., ECC scalar multiplication) \cite{steiner1996diffie}.  \textbf{(vi) OPE}: we rely on \ope~scheme proposed by Boldyreva~\cite{boldyreva2009order} for our evaluation because of its popularity and public implementation but our schemes can use {\em {any secure}} \ope~scheme (e.g.,~\cite{boldyreva2009order,popa2013ideal,kerschbaum2014optimal}) as a building block. \textbf{(vi) \En}: We rely on \aes~\cite{daemen1999aes}\footnote{AES is a symmetric block cipher adopted by the U.S. government and known to be the strongest symmetric crypto algorithm.} as our (\En,\De) for our cost analysis.
}
\end{flushleft}


\end{table*}
\section{Performance Evaluation}
\label{sec:PerformanceAnalysis}

We now evaluate our proposed schemes, \ROLPOS~and \LPGW, by comparing \ROLPOS~to its predecessor \LPOS~\cite{grissa2015location}, \ECEG~and \PPSS~as these schemes are all designed for the sensing architecture without a gateway, and comparing \LPGW~to \PDAFT~as both are designed for the sensing architecture with a gateway.

\subsection{Existing Approaches: \PPSS, \ECEG, and \PDAFT}

\PPSS~\cite{li2012location} uses secret sharing and the Privacy Preserving Aggregation (PPA) process proposed in~\cite{shi2011privacy} to hide the content of specific sensing reports and uses dummy report injections to cope with the DLP attack.

In~\ECEG,~\su s encrypt their \rss s with \fc's \ECELG~public key. One of the nodes aggregates these ciphertexts including its own and then sends the aggregated result to \fc. The \fc~then decrypts the aggregated result with its \ECELG~private key and makes the final decision.

\PDAFT~\cite{chen2014pdaft} combines Paillier cryptosystem~\cite{paillier1999public} with Shamir's secret sharing~\cite{shamir1979share}, where a set of smart meters sense the consumption of different households, encrypt their reports using Paillier, then send them to a gateway. The gateway multiplies these reports and forwards the result to the control center, which selects a number of servers (among all servers) to cooperate in order to decrypt the aggregated result. \PDAFT~requires a dedicated gateway, just like \LPGW, to collect the encrypted data, and a minimum number of working servers in the control center to decrypt the aggregated result.

\subsection{Performance Analysis and Comparison}
We focus on communication and computational overheads. We consider the overhead incurred during the sensing operations but not that related to system initialization (e.g. key establishment), where most of the computation and communication is done offline.
%
%
We model the membership change events in the network as a random process \ran~that takes on $0$ and $1$, and whose average is $\sr$. $\ran=0$ means that no change occurred in the network and $\ran=1$ means that some \su s left/joined the sensing task. Let \chr(t)~be a function that models the average number of \su s that join the sensing at the current sensing period $t$.

We precise that our performance analysis is not based on a simulation but rather on measuring the computational and communication overhead involved in the cryptographic operations that we deployed, like \yme~protocol and \ope. This gives us an idea about how our schemes perform compared to existent approaches in terms of incurred overhead. The execution times of the different primitives and protocols were measured on a laptop running Ubuntu 14.10 with 8GB of RAM and a core M 1.3 GHz Intel processor, with cryptographic libraries MIRACL~\cite{miracl}, Crypto++~\cite{crypto++} and {\em Louismullie}'s Ruby implementation of \ope~\cite{opeRuby}. C++ implementations that we developed for the optimized \ECELG~and the \yme~schemes will be provided for public use. 

\noindent \textbf{Computational Overhead}: Table \ref{tab:Table2} provides an analytical computational overhead comparison including the details of variables, parameters and the overhead of building blocks.

In \ROLPOS, \fc~requires only a logarithmic number of \yme~executions. An \su~requires a small constant number of {\em Point additions} $PAddQ$, one \ope~encryption and group key update, which is necessary only \sr~percent of the time when there is a change in the network (with only a logarithmic overhead in the number of \su s). The signature verification operation, that new \su s have to perform upon joining the sensing, is extremely fast in most of the digital signature schemes compared to the system overall computational overhead that we study in this section. This makes the delay introduced by the digital signature negligible compared to the overall computational overhead inferred by \ROLPOS~regardless of the used digital signature scheme. Thus, we don't consider this delay in our evaluation. This makes \ROLPOS~much more efficient than \ECEG~and \PPSS, especially for a relatively large number of \su s. 

In \LPGW, \fc~requires only a small constant number of $(\De,\En,\ope)$ operations. An \su~requires one \ope~and \En~encryptions of its \rss. Finally, \gw~requires one \De~operation per \su~and one \En~of vector $\mbox{\boldmath$\bin$}$. All computations in \LPGW~rely on only symmetric cryptography, which makes it {\em the most computationally efficient scheme among all alternatives} as discussed below.

\begin{figure}[h!]
  \centering
  \subfigure[Schemes w/o gateway]{\includegraphics[height=4.1cm,width=4.35cm]{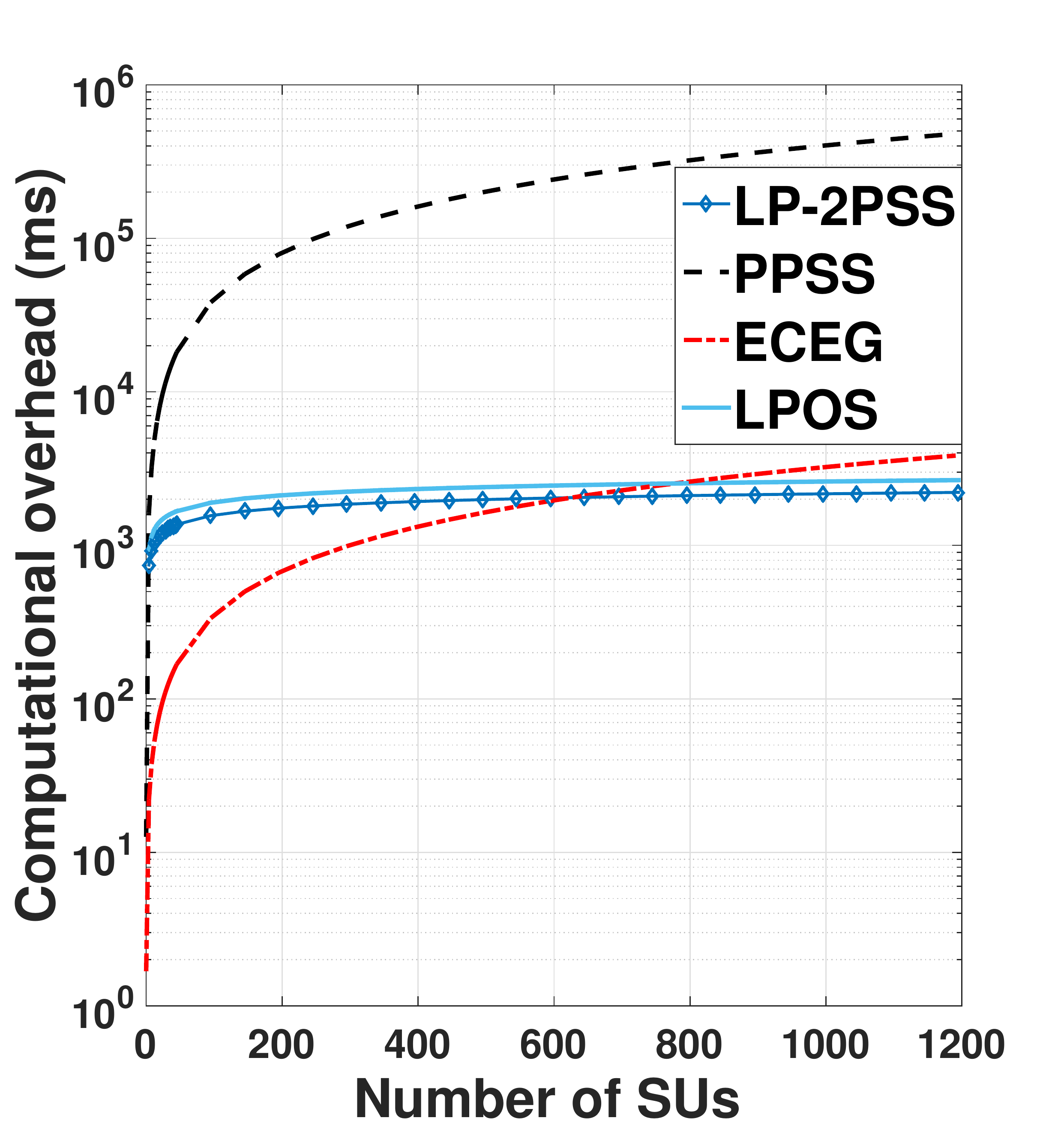}\label{fig:comp_overhead_wo_gw}}
  \subfigure[Schemes w/ gateway]{\includegraphics[height=4.1cm,width=4.35cm]{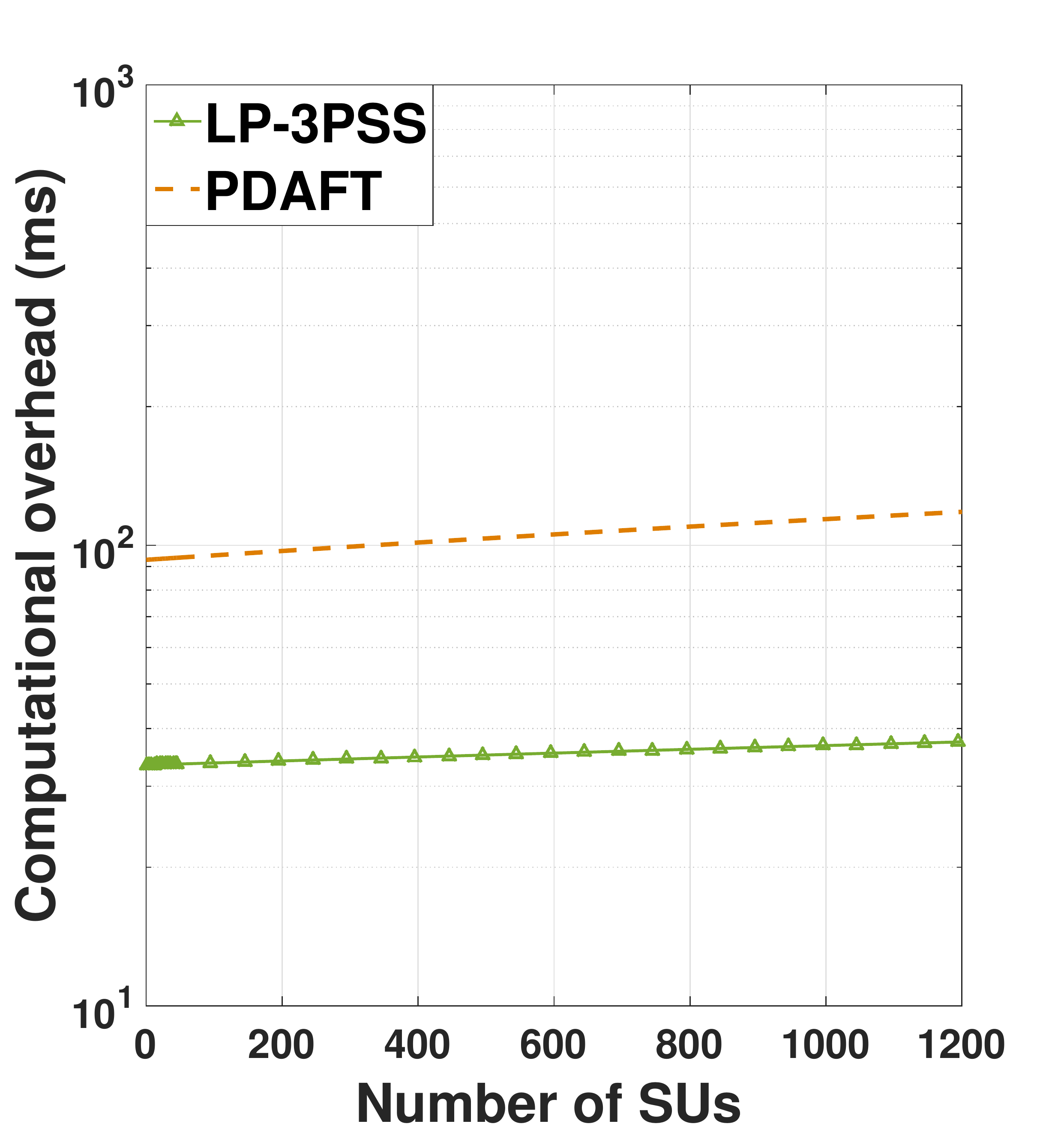}\label{fig:comp_overhead_w_gw}}
  \caption{Computation Overhead, $\chr = 5$, $\sr=20\%$ \& $\kap =80$} \label{fig:perfComp}
\end{figure}

For illustration purpose, we plot in Figure~\ref{fig:perfComp} the system end-to-end computational overhead of the different schemes.
Figure~\ref{fig:comp_overhead_wo_gw} shows that \ROLPOS~incurs an overhead that is comparable to that incurred by \ECEG, but much lower than that incurred by \PPSS. Figure~\ref{fig:comp_overhead_wo_gw} shows also that \ROLPOS~performs slightly better than its predecessor \LPOS.

Figure~\ref{fig:comp_overhead_w_gw} shows that \LPGW~is several order of magnitudes faster than \PDAFT~for any number of \su s.

 \begin{figure*}[h!]
  \centering
  \subfigure[FC: w/o gateway]{\includegraphics[scale=0.27]{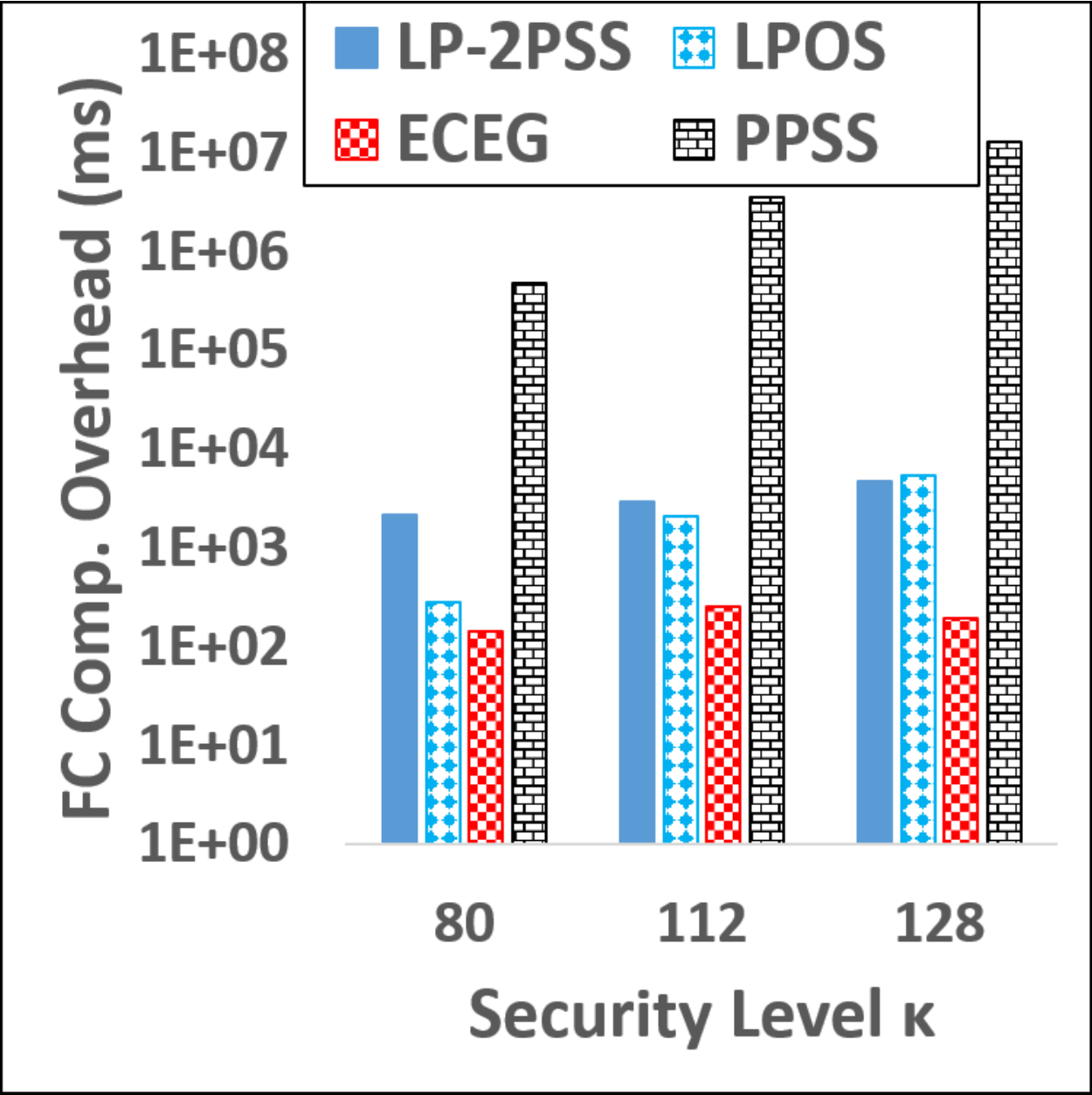}\label{fig:CompBarFCWOGW}}
\subfigure[FC: w/ gateway]{\includegraphics[scale=0.27]{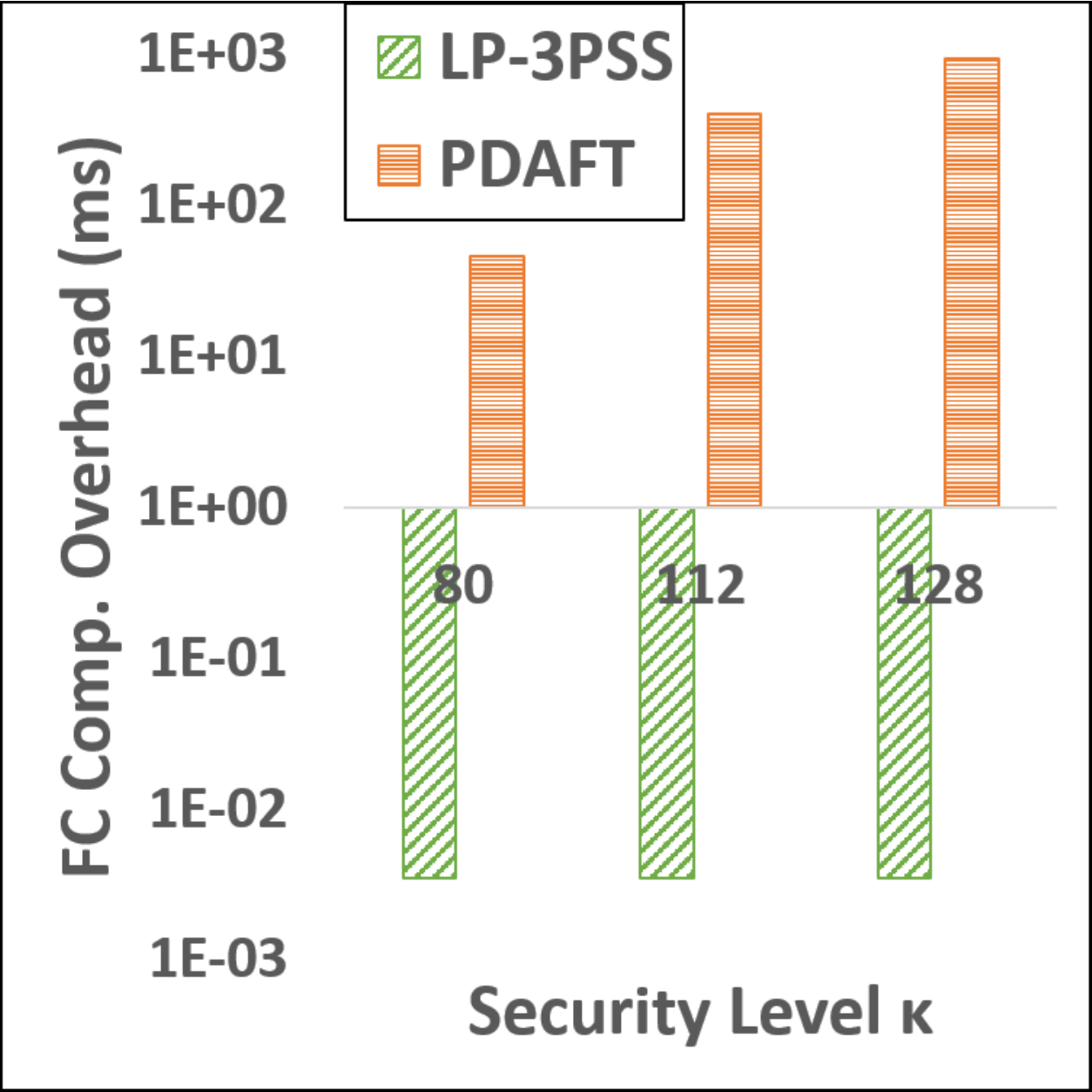}\label{fig:CompBarFCWG}}
  \subfigure[SU: w/o gateway]{\includegraphics[scale=0.27]{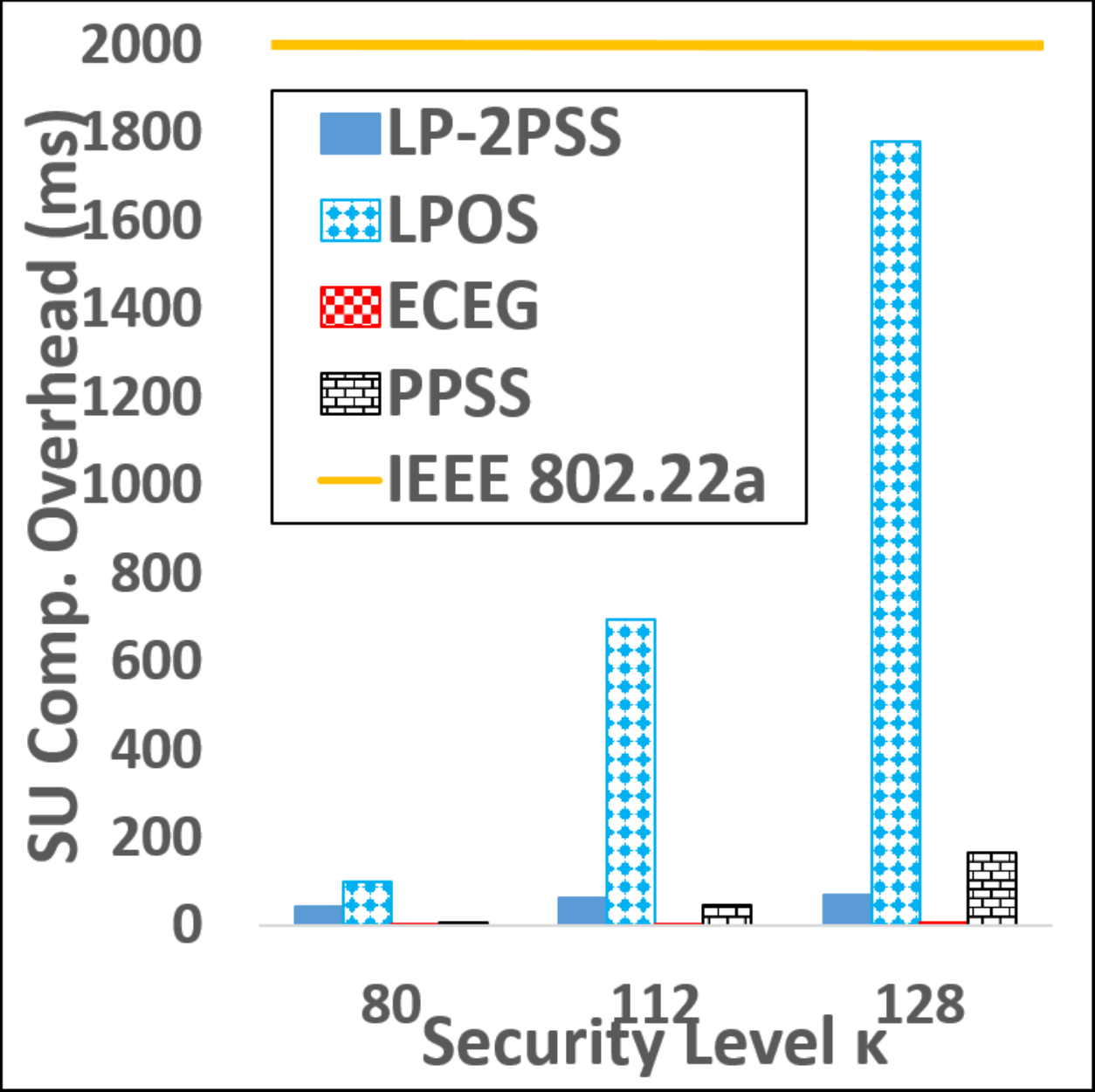}\label{fig:CompBarSUWOGW}}
  \subfigure[SU: w/ gateway]{\includegraphics[scale=0.27]{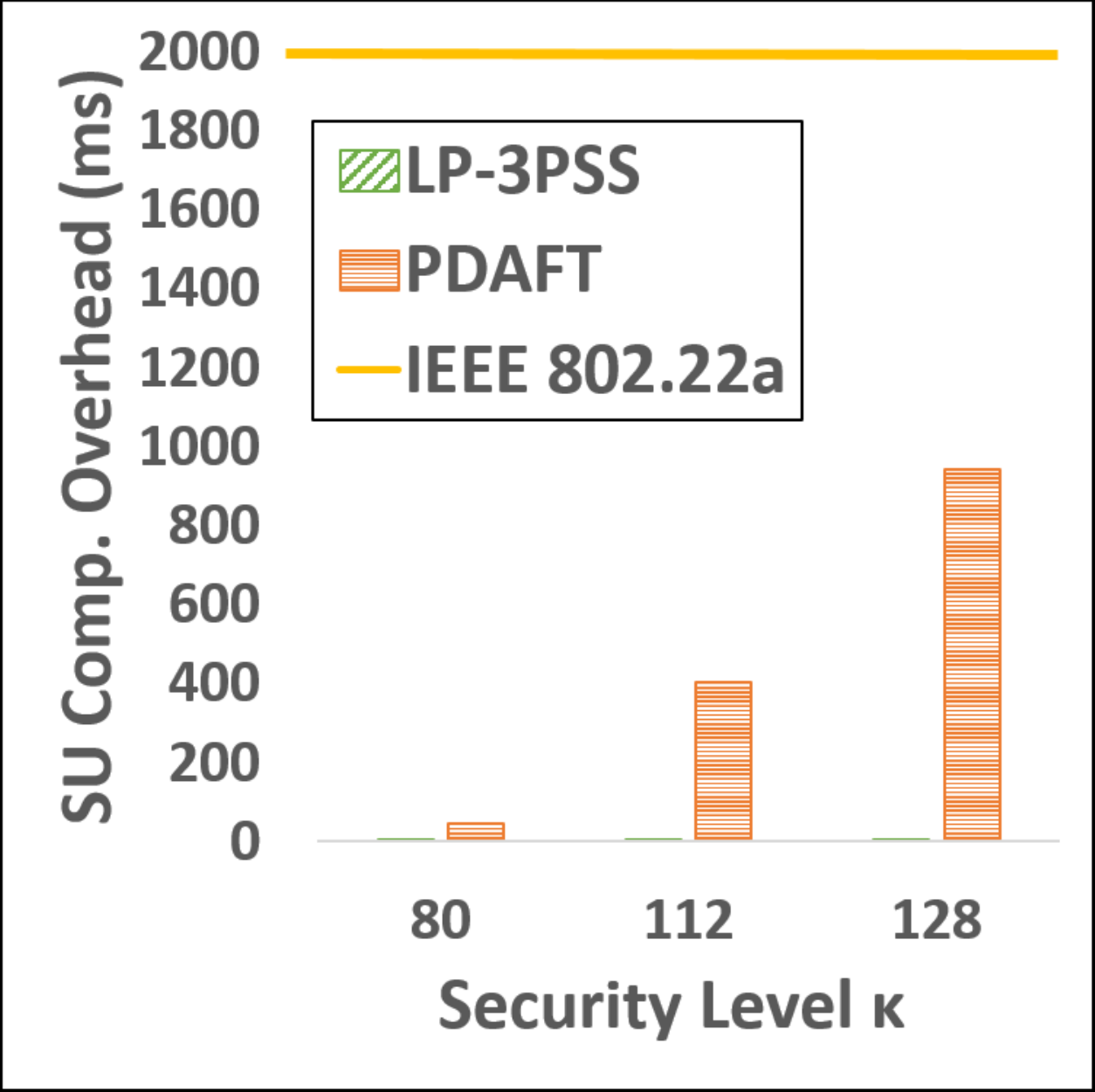}\label{fig:CompBarSUGW}}
  \subfigure[GW]{\includegraphics[scale=0.27]{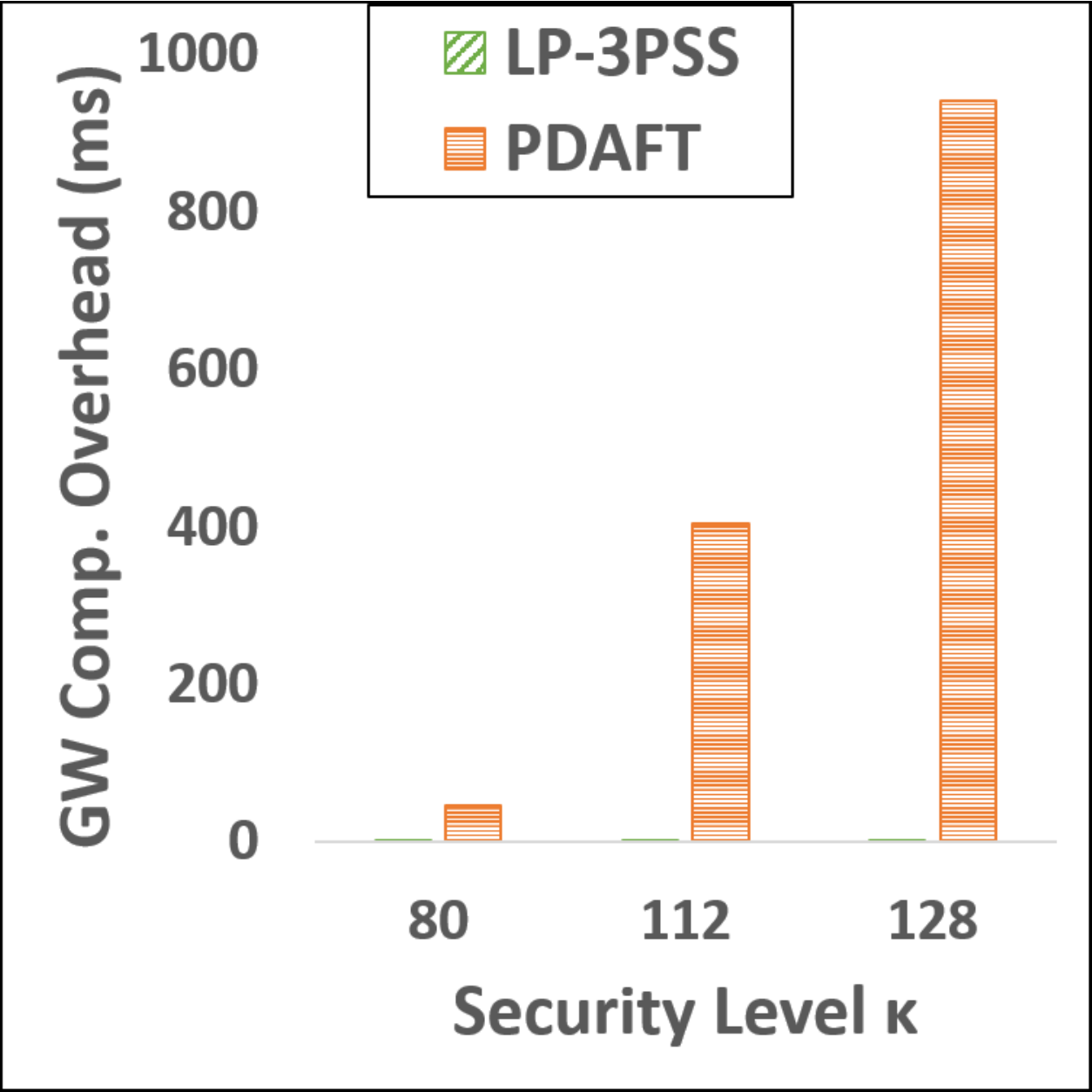}\label{fig:CompBarGW}}
  \caption{Computational Overhead Variation with Respect to Security Parameter $\kappa$ for $\nbr=1200$, $\chr = 5$ \& $\sr=20\%$} \label{fig:perfCompBar}
  \vspace{-0.1in}
\end{figure*}

Notice that the key generation and signing operations are done only once at the beginning of the protocol as \ta, and thus $T$, should be static over time unless a dramatic change in the system environment occurs which leads to the re-execution of the \ROLPOS's initialization phase. That is why these operations are not counted for the operational overhead of our scheme.

We also study the impact of the security parameter, $\kappa$, which controls the encryption key length, by varying it in accordance with {\em NIST}'s recommendations~\cite{keylength}. Note that this assesses the suitability of a scheme for a long term deployment in a stable networking infrastructure. Figure~\ref{fig:perfCompBar}, evaluating the schemes under three values of $\kappa$, shows that our schemes are the least impacted by increasing security parameters. It also shows that \LPGW~is significantly more efficient than \PDAFT~in terms of computation overhead for all entities. Note that our schemes achieve a delay, which is well below the $2$-second computation delay required by {\em IEEE 802.22 standard} for TV white space management~\cite{IEEEStd80222a}. This standard requires that the system handles dynamism in the network and that \rss~values lie within the interval $[-104,23.5]dB$ and are encoded under 8 bits. Figures~\ref{fig:CompBarFCWOGW} \& \ref{fig:CompBarSUWOGW} show the gain in computational performance of \ROLPOS~over \LPOS~especially for high security levels and from the \su s side.

\noindent \textbf{Communication Overhead}:
Table~\ref{tab:Table3} provides the analytical communication overhead comparison. \ROLPOS~requires $\log(\nbr)$ message exchanges for \ym~protocol, \nbr~\ope~ciphertexts and $\log(\nbr)$ messages for group key update (only needed \sr~percent of the time when there is a membership change). If some \su s join the \crn,  \ROLPOS~requires sharing the digital signature $\sign$ of message $T$ and the public key $\pkhors$ used to construct this signature with \chr~new \su s. \LPGW~requires (\nbr+1) \En~ciphertexts and single \ciphGW, which are significantly smaller than the values transmitted by \PDAFT. 

 \begin{table*}[h!]
 \scriptsize

\centering  \caption{Communication overhead comparison} \label{tab:Table3}
\renewcommand{\arraystretch}{1.25}{
\begin{tabular}{||c||c||}

\hline \multicolumn{1}{||c||}{\textbf{\em Scheme}}   &  \multicolumn{1}{|c||}{\textbf{Communication}} \\


\hline \hline \multicolumn{1}{||c||}{\textbf{\ROLPOS}} & $ 2\gam \cdot |Q| \cdot (2+log\:\nbr)+ \nbr  \cdot \epsilon_{\ope}+\sr\cdot|Q| \cdot log\:\nbr \;+ \;\chr\cdot(\vert \sign \vert + \vert \pkhors \vert)_{DS}$   \\
\hline \hline  \multicolumn{1}{||c||}{\textbf{\LPOS}} & $ 2\gam \cdot |p| \cdot (2+log\:\nbr)+ \nbr \cdot \epsilon_{\ope}+\sr\cdot|Q| \cdot log\: \nbr$  \\
\hline \multicolumn{1}{||c||}{\textbf{\ECEG}} & $4|Q| \cdot (4\nbr+\chr)$  \\
\hline \multicolumn{1}{||c||}{\PPSS} & $ |p|\cdot \nbr + \chr\cdot\sr\cdot |p| \cdot \nbr$  \\

\hline  \hline \multicolumn{1}{||c||}{\textbf{\LPGW}} & $ (\nbr + 1 )\cdot \blck$ \\
\hline \multicolumn{1}{||c||}{\PDAFT} & $|\rsa|\cdot (2(\nbr+1)+\chr)$  \\ \hline

\end{tabular}}
\begin{flushleft}
\scriptsize{
$\epsilon_{\ope} = 128\:bits$: maximum ciphertext size obtained under \ope~encryption, \blck: size of ciphertext under \En. $\vert \sign \vert$ and $\vert \pkhors \vert$ are respectively the size of the digital signature and the public key of the digital signature scheme $DS$.
}
\end{flushleft}
\vspace{-3mm}
\end{table*}

\begin{figure}[h!]
  \centering
  \subfigure[Schemes w/o. gateway]{\includegraphics[height=4.1cm,width=4.35cm]{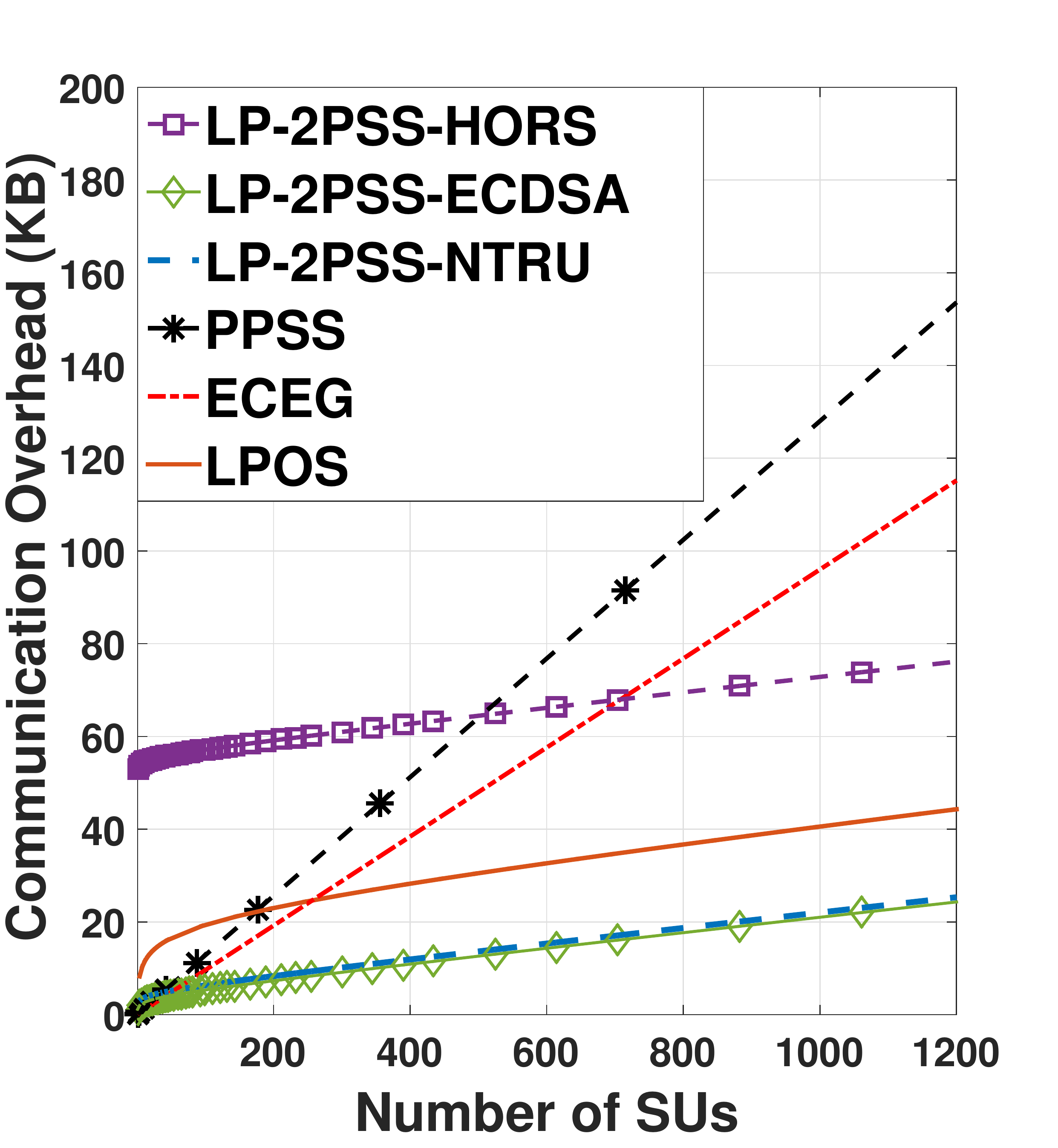}\label{fig:comm_overhead_wo_gw}}
  \subfigure[Schemes w. gateway]{\includegraphics[,height=4.1cm,width=4.35cm]{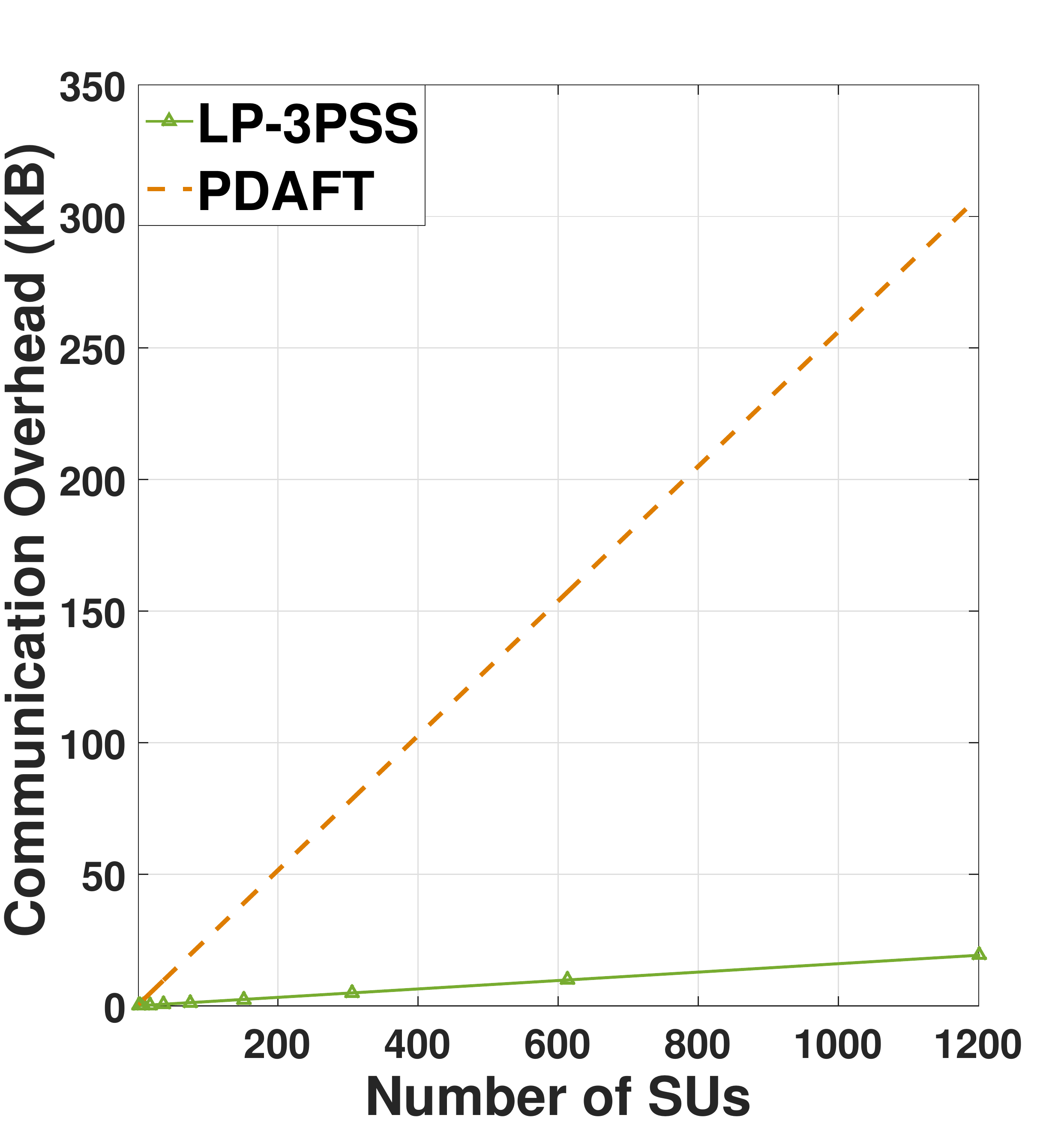}\label{fig:comm_overhead_w_gw}}
  \caption{Communication Overhead, $\chr = 5$, $\sr=20\%$ \& $\kap =80$} \label{fig:perfComm}
\end{figure}

We further compare our schemes with their counterparts in terms of communication overhead in Figure~\ref{fig:perfComm}. Figure~\ref{fig:comm_overhead_wo_gw} illustrates the communication overhead induced by \ROLPOS~using different digital signature schemes (\HORS, {\em ECDSA} and {\em NTRU}) compared to the original scheme, \LPOS, and also to existent approaches \PPSS~and \ECEG. This Figure shows that \ROLPOS~is more efficient than \PPSS~and \ECEG~due to the use of elliptic curve cryptography with smaller key sizes. Using {\em ECDSA} or {\em NTRU} seems to be the best option in terms of communication overhead as expected. 
For a large number of \su s, using a digital signature scheme with large signature size like \HORS~does not prevent \ROLPOS-\HORS~from performing better than existent approaches especially for a large number of \su s. Figure~\ref{fig:comm_overhead_w_gw} shows that \LPGW~has the smallest communication overhead 
when compared with \PDAFT, since it relies on symmetric cryptography only.  \PPSS~and \PDAFT~have a very high communication overhead due to the use of expensive public key encryptions (e.g., Pailler~\cite{paillier1999public}).

\begin{figure}[h!]
  \centering

  \subfigure[Schemes w/o. gateway]{\includegraphics[scale=0.27,height=3.3cm]{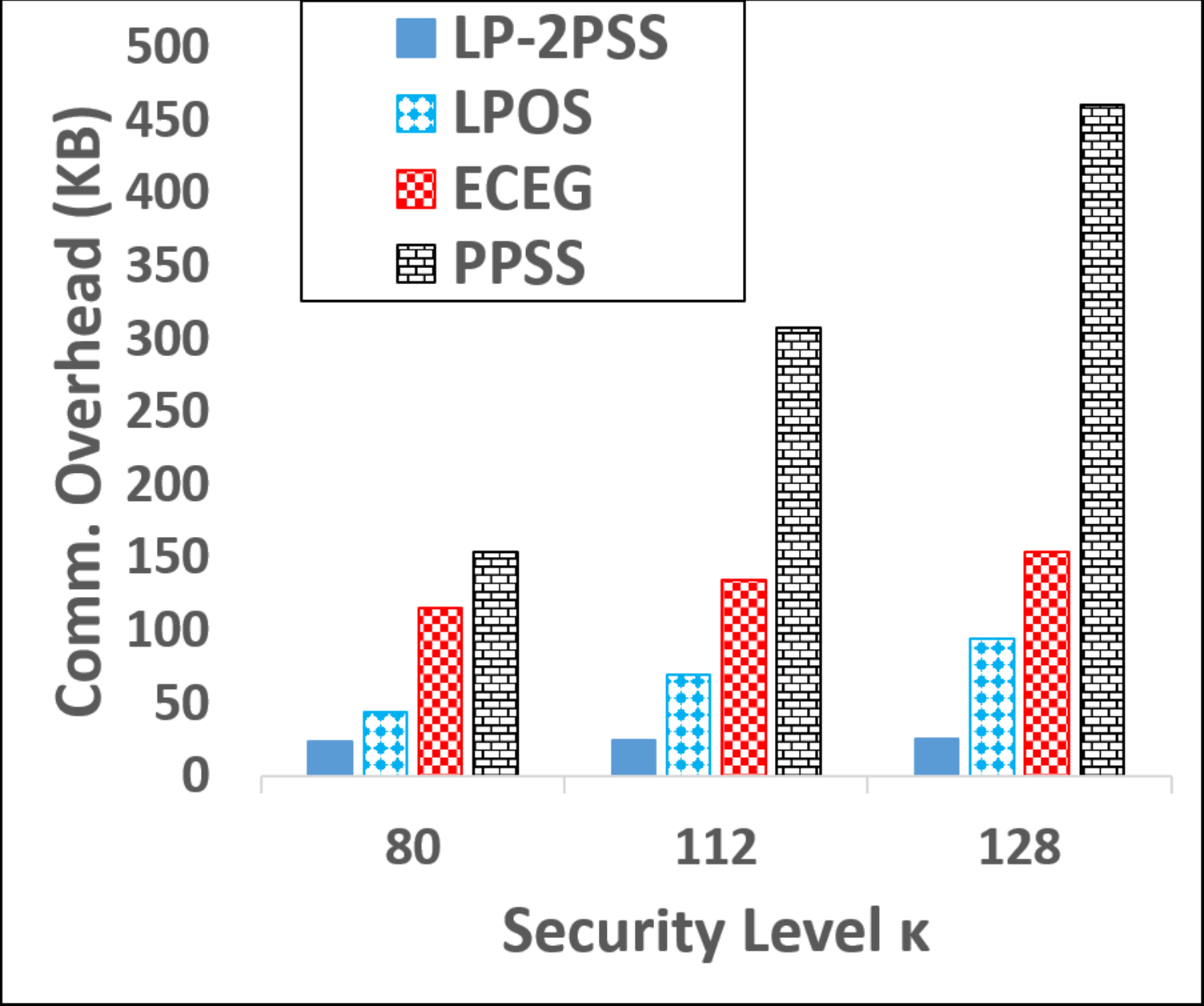}\label{fig:commBarWOGW}}
  \subfigure[Schemes w. gateway]{\includegraphics[scale=0.27,height=3.3cm]{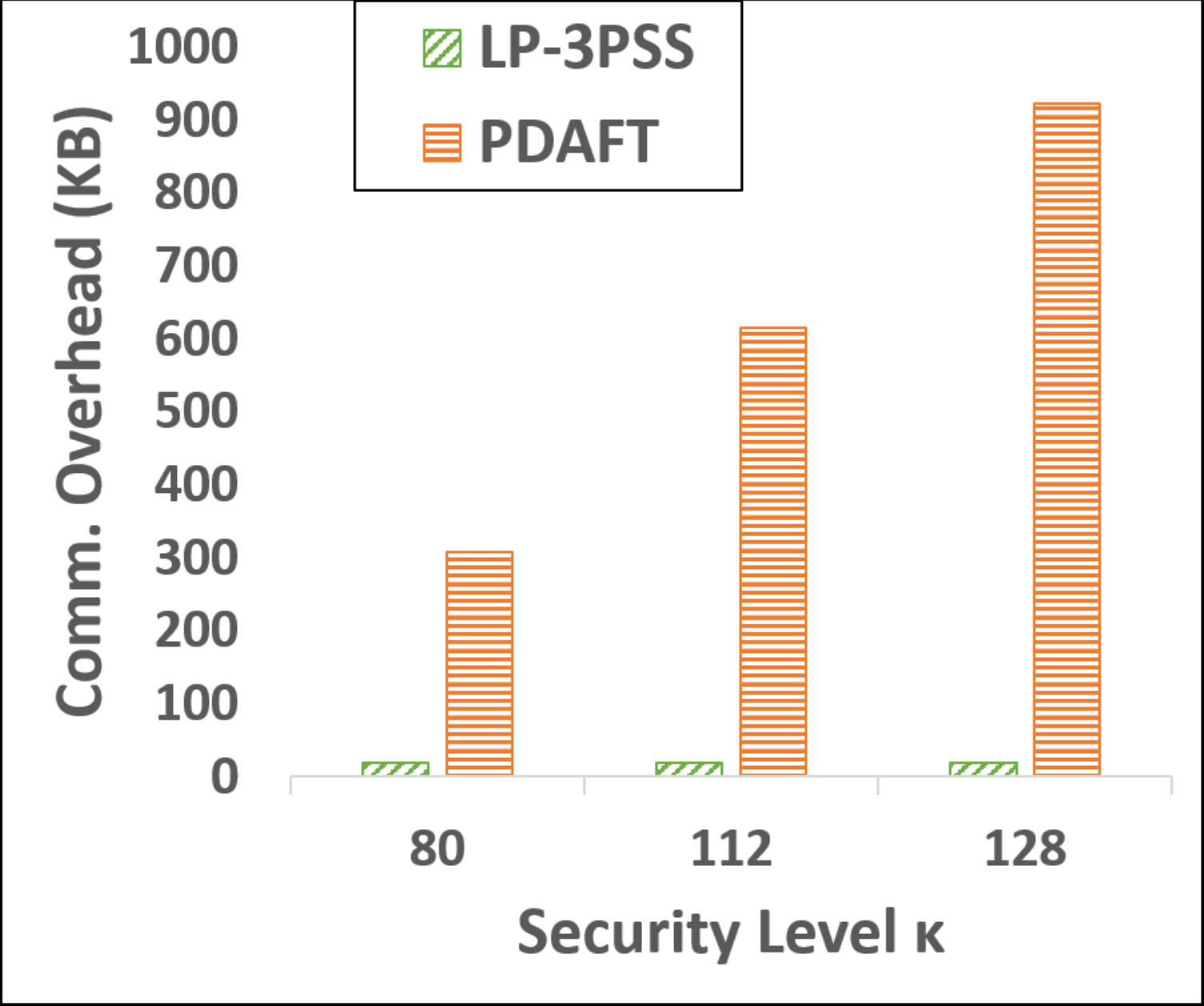}\label{fig:commBarGW}}
  \caption{Communication Overhead with varying $\kappa=(80,112,$ $128)$ for $\nbr=1200$, $\chr = 5$ \& $\sr=20\%$.} \label{fig:perfCommBar}
  \vspace{-5mm}
\end{figure}

We also study and show in Figure~\ref{fig:perfCommBar} the impact of the security parameter, $\kappa$, on the communication overhead. Note that the performance gap between our schemes and their counterparts drastically grows when $\kappa$ is increased, showing the suitability of our schemes for long term deployment. Our schemes possess this desirable feature, thanks to their innovative use of compact cryptographic primitives.
Figures~\ref{fig:comm_overhead_wo_gw} \& \ref{fig:commBarWOGW} show again how efficient \ROLPOS~is compared to the original \LPOS~in terms of communication overhead.

Overall, our performance analysis indicates that \LPGW~is more efficient than \ROLPOS, and significantly more efficient than all other counterpart schemes in terms of computation and communication overhead, even for increased values of the security parameters, but with the cost of including an additional entity. Moreover, Figures~\ref{fig:perfCompBar} \& \ref{fig:perfCommBar} show that our schemes are impacted much less by increased security parameters when compared to existing alternatives, and therefore are ideal for long term deployment. Note that our performance analysis lacks the evaluation of the {\em SCPU}-based version of \LPGW~due to the fact that this hardware is very expensive.


\section{Conclusion}
\label{sec:Conclusion}
We developed two efficient schemes for cooperative spectrum sensing that protect the location privacy of \su s with a low cryptographic overhead while guaranteeing an efficient spectrum sensing. Our schemes are secure and robust against \su s' dynamism, failures, and maliciousness. Our performance analysis indicates that our schemes outperform existing alternatives in various metrics. 


\ifCLASSOPTIONcaptionsoff
  \newpage
\fi


\small{
\bibliographystyle{IEEEtran}
\bibliography{IEEEabrv,references}
}

%

\begin{IEEEbiography}[{\includegraphics[width=1in,height=1.25in,clip,keepaspectratio]{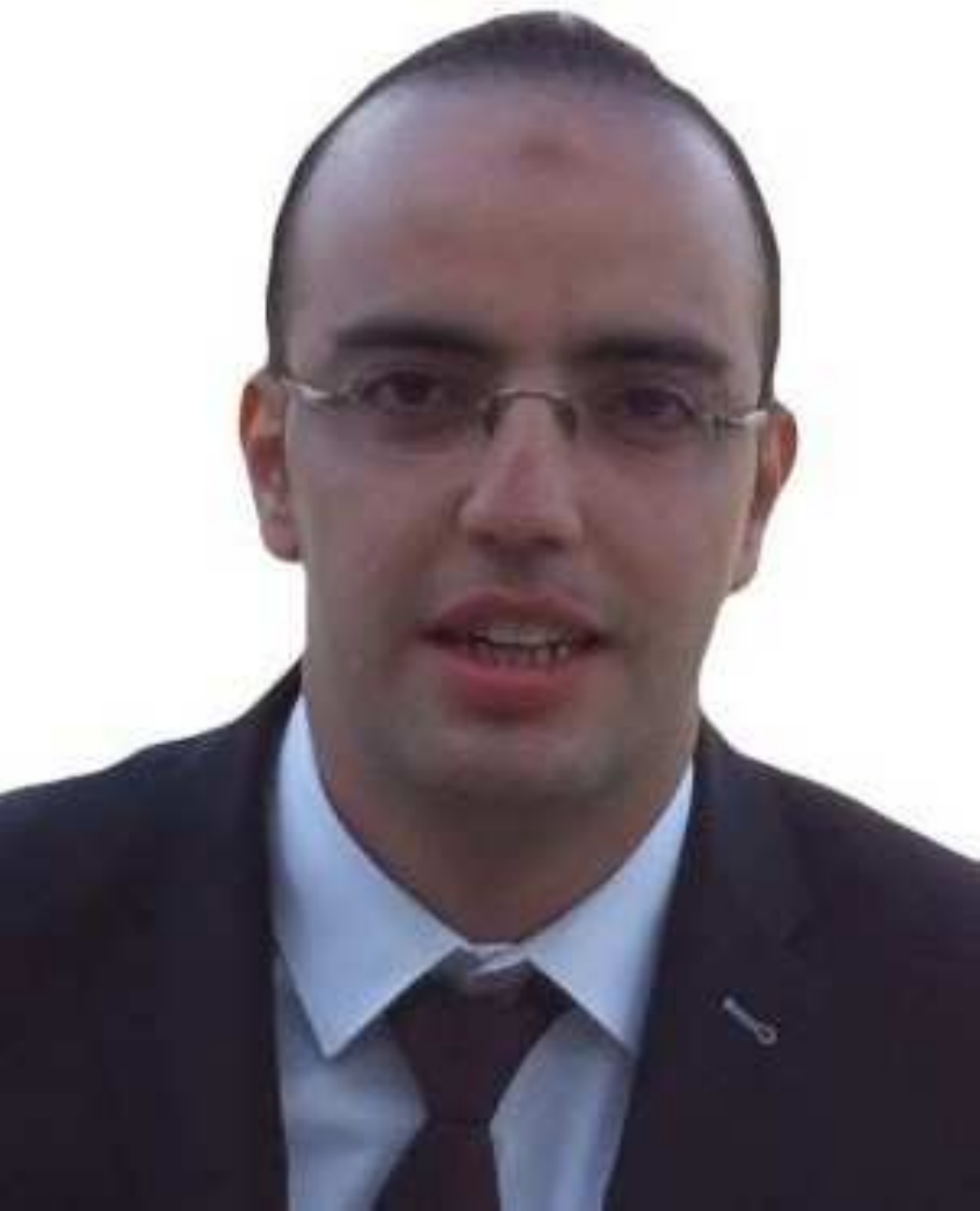}}]{Mohamed Grissa}(S'14) received the Diploma of Engineering (with highest distinction)
in telecommunication engineering from Ecole
Superieure des Communications de Tunis, Tunis,
Tunisia, in 2011, and the
M.S. degree in electrical and computer engineering
(ECE) from Oregon State University, Corvallis, OR, USA, in 2015. He is currently working
toward the Ph.D. degree at the School of Electrical
Engineering and Computer Science (EECS), Oregon
State University, Corvallis, OR, USA.

Before pursuing the Ph.D. degree, he worked as a Value Added Services Engineer at Orange France Telecom Group from 2012 to 2013. His research interests include privacy and security in wireless networks, cognitive radio networks, IoT and eHealth systems.
\end{IEEEbiography}

\begin{IEEEbiography}[{\includegraphics[width=1in,height=1.25in,clip,keepaspectratio]{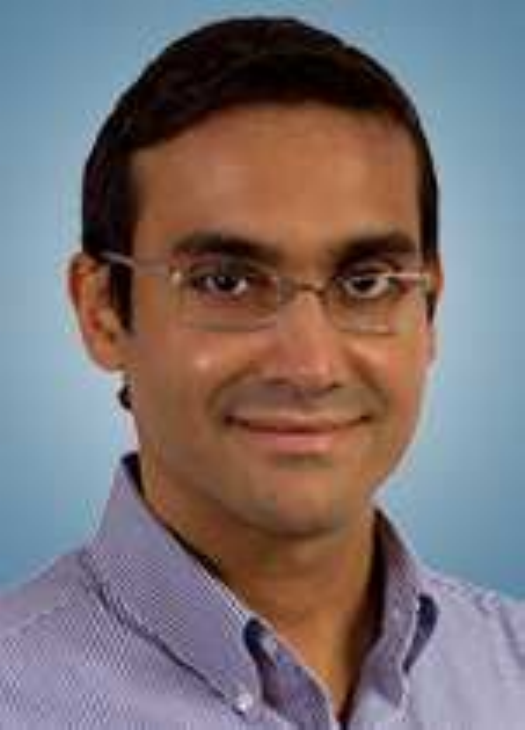}}]{Attila A. Yavuz} (S'05\textendash M'10) received a BS degree in Computer Engineering from Yildiz Technical University (2004) and a MS degree in Computer Science from Bogazici University (2006), both in Istanbul, Turkey. He received his PhD degree in Computer Science from North Carolina State University in August 2011. Between December 2011 and July 2014, he was a member of the security and privacy research group at the Robert Bosch Research and Technology Center North America. Since August 2014, he has been an Assistant Professor in the School of Electrical
Engineering and Computer Science, Oregon State University, Corvallis, USA. He is also an adjunct faculty at the University of Pittsburgh's School of Information Sciences since January 2013.
 
Attila A. Yavuz is interested in design, analysis and application of cryptographic tools and protocols to enhance the security of computer networks and systems. His current research focuses on the following topics: Privacy enhancing technologies (e.g., dynamic symmetric and public key based searchable encryption), security in cloud computing, authentication and integrity mechanisms for resource-constrained devices and large-distributed systems, efficient cryptographic protocols for wireless sensor networks.
\end{IEEEbiography}
%
%
\begin{IEEEbiography}[{\includegraphics[width=1in,height=1.25in,clip,keepaspectratio]{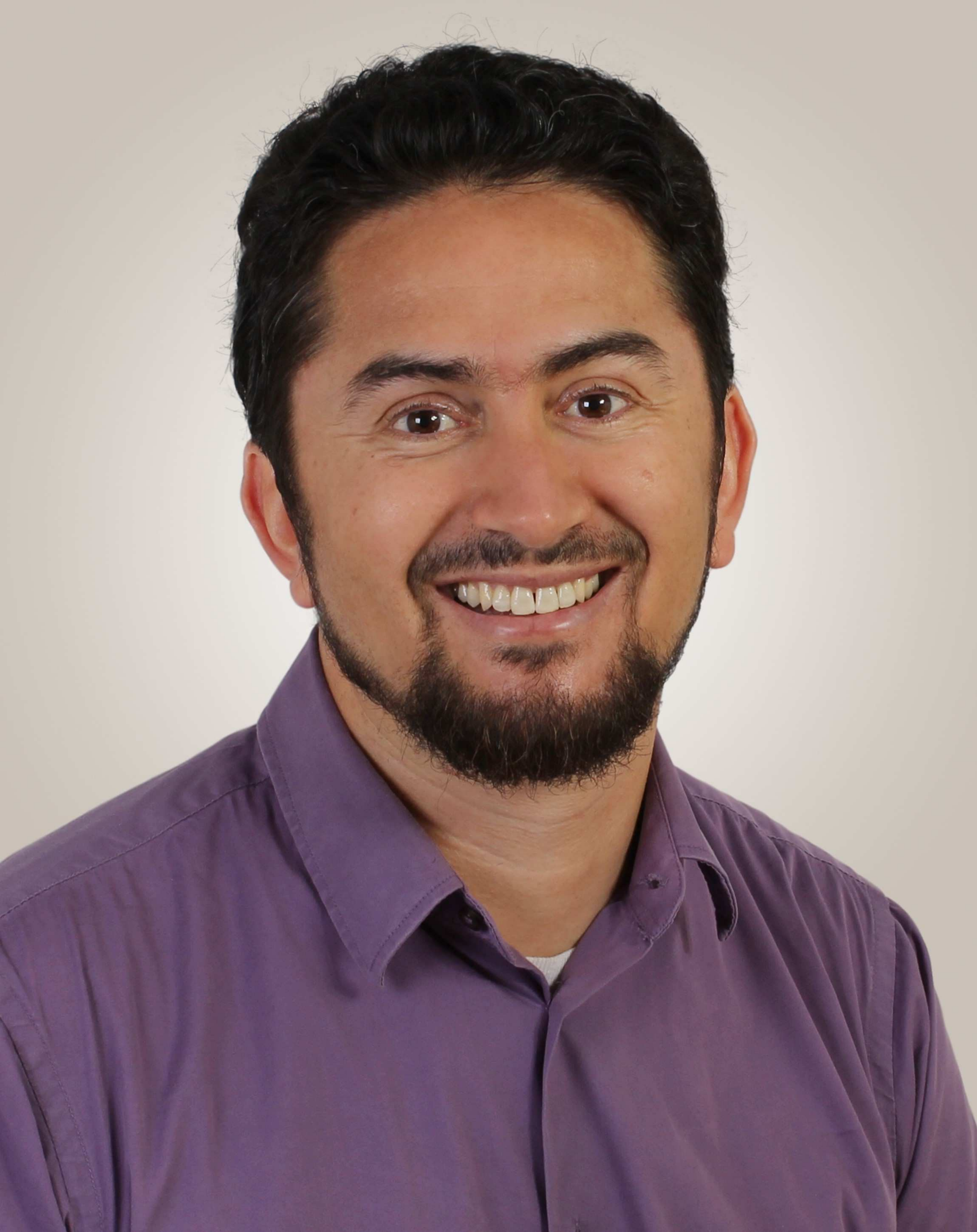}}]{Bechir Hamdaoui} (S'02\textendash M'05\textendash SM'12) is presently an Associate Professor in the School of EECS at Oregon State University. He received the Diploma of Graduate Engineer (1997) from the National School of Engineers at Tunis, Tunisia. He also received M.S. degrees in both ECE (2002) and CS (2004), and the Ph.D. degree in Computer Engineering (2005) all from the University of Wisconsin-Madison. His current research focus is on distributed resource optimization, parallel computing, cognitive computing \&~networking, cloud computing, and Internet of Things. He has won the NSF CAREER Award (2009), and is presently an AE for IEEE Transactions on Wireless Communications (2013-present), and Wireless Communications and Mobile Computing Journal (2009-present). He also served as an AE for IEEE Transactions on Vehicular Technology (2009-2014) and for Journal of Computer Systems, Networks, and Communications (2007-2009). He is currently serving as the chair for the 2016 IEEE INFOCOM Demo/Posters program and the chair for the 2016 IEEE GLOBECOM Wireless and Mobile Networks symposium. He has also served as the chair for the 2011 ACM MOBICOM’s SRC program, and as the program chair/co-chair of several IEEE symposia and workshops (including ICC 2014, IWCMC 2009-2016, CTS 2012, PERCOM 2009). He also served on technical program committees of many IEEE/ACM conferences, including INFOCOM, ICC, GLOBECOM, and others. He has been selected as a Distinguished Lecturer for the IEEE Communication Society for 2016 and 2017. He is a Senior Member of IEEE, IEEE Computer Society, IEEE Communications Society, and IEEE Vehicular Technology Society.
\end{IEEEbiography}

\vfill


\end{document}